\documentclass[a4paper,usenatbib,useAMS]{mn2e}

\usepackage{epsf}
\usepackage[dvips]{epsfig}
\usepackage{times}
\usepackage{wasysym}

\begin{document}

\label{firstpage}

\title[mass loss of UCDs]{Mass loss and expansion of
  ultra compact dwarf galaxies through gas expulsion and stellar evolution for top-heavy stellar initial mass functions}

\author[Dabringhausen et al.]{
J. Dabringhausen$^{1}$ \thanks{E-mail: joedab@astro.uni-bonn.de},
M. Fellhauer$^{2}$ \thanks{mfellhauer@astro-udec.cl} and
P. Kroupa$^{1}$ \thanks{pavel@astro.uni-bonn.de}\\
$^{1}$ Argelander-Institut f\"ur Astronomie, Universit\"at Bonn, Auf dem
H\"ugel 73, 53121 Bonn, Germany\\
$^{2}$ Departamento de Astronomia, Universidad de Concepcion, Casilla 160-C, Concepcion, Chile}

\pagerange{\pageref{firstpage}--\pageref{lastpage}} \pubyear{2009}

\maketitle

\begin{abstract}
The dynamical $V$-band mass-to-light ratios of ultra compact dwarf galaxies (UCDs) are higher than predicted by simple stellar population models with the canonical stellar initial mass function (IMF). One way to explain this finding is a top-heavy IMF, so that the unseen mass is provided by additional remnants of high-mass stars.  A possible explanation for why the IMF in UCDs could be top-heavy while this is not the case in less massive stellar systems is that  encounters between proto-stars and stars become probable in forming massive systems. However, the required number of additional stellar remnants proves to be rather high, which raises the question of how their progenitors would affect the early evolution of a UCD. We have therefore calculated the first 200~Myr of the evolution of the UCDs, using the particle-mesh code Superbox. It is assumed that the stellar populations of UCDs were created in an initial starburst, which implies heavy mass loss during the following $\approx 40$ Myr due to primordial gas expulsion and supernova explosions. This mass loss is modelled by reducing the mass of the particles according to tabulated mass loss histories which account for different IMFs, star formation efficiencies (SFEs), heating efficiencies (HEs), initial masses and initial extensions of the computed UCDs. For each combination of SFE and HE we find objects that roughly resemble UCDs at the end of the simulation. For low SFEs, the IMF would have to be steeper than in the case of very high SFEs for the models not to expand too much. However, the main conclusion is that the existence of UCDs does not contradict the notion that their stellar populations formed rapidly and with a top-heavy IMF. We find tentative evidence that the UCDs may have had densities as high as $10^8 \rm{M}_{\odot} \, \rm{pc}^{-3}$ at birth. This will have to be confirmed by follow-up modelling.
\end{abstract}

\begin{keywords}
  galaxies: evolution -- galaxies: dwarfs -- galaxies: star clusters -- methods: N-body simulations
\end{keywords}

\section{Introduction}
\label{sec:intro}

Ultra compact dwarf galaxies (UCDs) are stellar systems with total stellar masses between $10^6$ and $10^8 \rm{M}_{\odot}$ and projected half-light radii of $\apprle 50 \, \rm{pc}$ \citep{Hil1999, Dri2000, Dri2003, Phi2001, Has2005}. They can be considered to be galaxies because of their high median two-body relaxation times, $t_{\rm{rh}}$, which are at least of the order of a Hubble time, $\tau_{\rm{H}}$, while star clusters, including globular clusters (GCs), have $t_{\rm{rh}}<\tau_{\rm{H}}$ \citep{Kro1998,Dab2008}.

One of the most intriguing properties of UCDs are their generally high dynamical $M/L_V$ ratios \citep{Dab2008,Mie2008b}. Different explanations have been suggested for this finding, such as the presence of non-baryonic cold dark matter (CDM) in them (e.g. \citealt{Has2005} and \citealt{Goe2008}) or the disturbance of UCDs by the tidal field of a massive galaxy \citep{Fel2006}. However, if dwarf spheroidal galaxies (dSphs) are indeed DM dominated\footnote{There is an ongoing debate on the origin of the dSphs around the Milky Way. Their disk-like distribution has a natural explanation if the dSphs are ancient tidal dwarf galaxies instead of DM-dominated primordial galaxies (\citealt{Met2009} and references therein). The high $M/L$ ratios derived for them would in this scenario either be the consequence of the assumption of virial equilibrium not holding for them \citep{Kro1997} or would imply that Newtonian gravity cannot be applied in the limit of very weak fields. A tidal origin of dSphs may suggest the same for dwarf elliptical galaxies, since \citet{Kor1985} argues that these two populations may actually be the same type of galaxies.} and if UCDs are located at the centre of the same type of haloes as dSphs, the DM-density in UCDs would be two orders of magnitude too low to explain their elevated $M/L_V$ ratios, although adiabatic contraction \citep{Blu1986} may alleviate this problem \citep{Mur2008}. Tidal distortion can explain the high $M/L_V$ ratios of only a few UCDs out of a larger sample, as it requires quite specific orbital parameters in order to have an observable effect. On the other hand, the massive star cluster W3 in the merger remnant galaxy NGC 7252 has a mass and a projected half-light radius typical for a UCD, while its age suggests that it formed during the merger of the progenitors of NGC 7252 \citep{Mar2004}. \citet{Fel2005b} have shown that star cluster complexes as observed in interacting sytems like the Antennae (NGC 4038 and NGC 4039) are likely to evolve into an object similar to W3 on the required time-scale, but stellar systems originating from tidal interactions would essentially be CDM-free \citep{Her1992}. In summary, an unusual stellar initial mass function (IMF) appears to be an attractive and physically plausible alternative for explaining the $M/L$ ratios of UCDs.

The IMF is a function defining the mass spectrum of stars born in a single star-formation event. If age, metallicity and IMF of a stellar population are known, its $M/L_V$ ratio can be calculated. For a given metallicity and a high enough age, a high $M/L_V$ ratio of a stellar population would either indicate an IMF with very many low-mass stars (bottom-heavy IMF) or an IMF with very many high-mass stars (top-heavy IMF). In the case of a top-heavy IMF, the high $M/L_V$ of the stellar population is the consequence of a high number of stellar remnants, which contribute mass, but almost no $V$-band luminosity. As an explanation for the $M/L_V$ ratios in UCDs, a bottom-heavy IMF has been discussed in \citet{Mie2008}, while a top-heavy IMF has been discussed in \citet{Mur2008} and Dabringhausen, Kroupa \& Baumgardt (2009), hereafter DKB.

Proposing a variability of the IMF might seem daring at first sight, because so far surveys of stars have failed in providing supportive evidence for this notion \citep{Kro2001,Kro2002,Kum2008}. This finding implies an invariant, universal IMF, which is referred to as the canonical IMF. It can be written as
\begin{equation}
\xi_{\rm{c}}(m_*) =k k_i m_*^{-\alpha_i},
\label{eq:IMF}
\end{equation}
with 
 
\begin{math}
\begin{array}{@{\hspace{-0.6cm}}lll}
&&\\[-4pt]
\alpha_1 = 1.3, & \  k_1=1, & \  0.1  \le  \frac{m_*}{\rm{M}_{\odot}} < 0.5,\\[3pt]
\alpha_2 = 2.3, & \  k_2=k_1\, 0.5^{\alpha_2-\alpha_1}=0.5, & \  0.5 \le  \frac{m_*}{\rm{M}_{\odot}} \le m_{\rm{max}},\\[-4pt]
&&\\
\end{array}
\end{math}
where $m_*$ is the initial stellar mass, $m_{\rm{max}}$ is the upper mass limit of the IMF, the factors $k_i$ ensure that the IMF is continuous where the power changes and $k$ is a normalisation constant \citep{Kro2008}. The subscript c identifies the canonical IMF. $\xi_{\rm{c}}(m_*)$ equals 0 if $m_*<0.1 \, \rm{M}_{\odot}$ or $m_*>m_{\rm max}$. For stellar systems as massive as the UCDs, $m_{\rm{max}}$ is equal to the maximum mass for stars, which is close to $150 \, \rm{M}_{\odot}$ \citep{Wei2004,Oey2005,Fig2005}. For any IMF, $dN=\xi(m_*)dm_*$ is the number of born stars in the mass interval $[m_*,m_*+dm_*]$. In the present paper, the constant $k$ is chosen such that
\begin{equation}
\int^{m_{\rm{max}}}_{0.1} \xi (m_*)m_* \, dm_*=1 \, \rm{M}_{\odot}.
\label{eq:norm1}
\end{equation}
Using this normalisation,
\begin{equation}
N=\int^{m_{\rm{max}}}_{0.1} \xi (m_*) \, dm_*
\label{eq:norm2}
\end{equation}
is formally the number of stars whose total mass is $1 \, \rm{M}_{\odot}$. Multipliying equation \ref{eq:norm2} by the factor $M_{*,0}/\rm{M}_{\odot}$ therefore equals the initial number of stars in a star cluster with an initial stellar mass of $M_{*,0}$ and the mean stellar mass, $\overline{m}$, equals equation (\ref{eq:norm1}) divided by equation (\ref{eq:norm2}).

Note that there are limitations to the determination of the IMF from star counts. For instance, massive stars are short-lived, which is why this approach can only give the high-mass IMF for recent star formation events. Low-mass stars on the other hand can be almost as old as the Universe, but they can only be detected very locally.

The existence of a universal law for the stellar mass spectrum would indeed be surprising from a theoretical point of view, since models for star-formation predict that the stellar mass spectrum depends on the conditions under which star formation takes place (e.g. \citealt{Ada1996}, \citealt{Mur1996}, \citealt{Lar1998} and \citealt{Cla2007}). Moreover, a top-heavy IMF is in fact required in a number of astrophysical models. This includes, besides the model proposed in DKB for the UCDs, also models for globular clusters (GCs) \citep{Dan2004,Pra2006,Dec2007}\footnote{If UCDs are indeed the most massive GCs, as proposed for instance in \citet{Mie2002}, \citet{Mie2004} and \citet{For2008}, it is evident that a top-heavy IMF in GCs suggests the same for UCDs. Note however that residual gas expulsion from mass-segregated clusters alleviates the need of a top-heavy IMF in GCs \citep{Dec2008}.}, distant galaxies \citep{Bau2005,Nag2005,Dok2008,Cha2008} and the Galactic centre \citep{Man2007}. The motivations for the top-heaviness of the IMF in these models include a higher ambient temperature at the time when the observed population formed and violent star formation in particularly dense gas. These conditions were likely to be given in the young UCDs, since the universe was much younger when they formed (i.e. the temperature of the cosmic microwave background was higher). Furthermore, the $\alpha$-enrichment found by \citet{Evs2007} in most of the Virgo-UCDs suggests rapid star-formation.

However, if stellar remnants are to account for the unseen mass in the UCDs, the top-heavyness of the IMF would have to be very pronounced. Introducing an IMF that equals the canonical IMF below $1 \, \rm{M}_{\odot}$ but has a different slope, $\alpha$, for $m>1 \, \rm{M}_{\odot}$, DKB suggest $1.0 < \alpha < 1.6$, depending on the age of the UCDs. These high-mass IMF slopes imply that the clear majority of the total initial stellar mass was locked up in stars more massive than $8 \, \rm{M}_{\odot}$, in contrast to the case with the canonical IMF. These stars have a very high luminosity and evolve rapidly, which makes their abundance a key issue for the evolution of a stellar system.

\begin{table*}
  \centering
  \caption{The initial parameters and some derived quantities for the UCD-models. The last two lines show the models for newly formed ONC-type star clusters from \citet{Kro2001a} for comparison. The columns denote the identification number of the model, the initial Plummer-radius $R_{\rm pl,0}$, its \emph{total} initial mass $M_{\rm pl,0}$, its \emph{stellar} initial mass $M_{*,0}$, the star formation efficiency (SFE$=M_{0,*}/M_{\rm{pl,0}}$), the heating efficiency, the initial characteristic crossing-time $T_{\rm cr}$, the initial characteristic three-dimensional velocity dispersion $\sigma_{\rm 3D,0}$, the initial central mass density and an estimate for the time-scale on which a given proto-star encounters a star during the formation of the UCD (see Section \ref{sec:encounters}).} 
  \label{tab:ini}
  \begin{tabular}{llllllllll} \hline
    model & $R_{\rm pl,0}$ & $M_{\rm pl,0}$ & $M_{*,0}$ & SFE & HE & $T_{\rm cr}$ &
    $\sigma_{\rm 3D,0}$ & $\rho_{\rm pl,0,c}$ & $t_{\rm{enc}}$\\
    & [pc] & [M$_{\odot}$] & [M$_{\odot}$] & & & [Myr] & [km\,s$^{-1}$] & [$10^6\rm{M}_{\odot} \, \rm{pc}^{-3}$] & Myr\\ \hline
    m7\_r3\_s1\_h1   	  & 3.0     & $1.0\times10^{7}$   & $1.0\times10^{7}$   & 1.0   & 1.0   & 0.153 & 65.0   & 0.088 &  0.23\\
    m7\_r5\_s1\_h1   	  & 5.0     & $1.0\times10^{7}$   & $1.0\times10^{7}$   & 1.0   & 1.0   & 0.330 & 50.3   & 0.019 & 1.4\\
    m8\_r5\_s1\_h1   	  & 5.0     & $1.0\times10^{8}$   & $1.0\times10^{8}$   & 1.0   & 1.0   & 0.104 & 159.0 & 0.191 & 0.043\\ \hline
    m7\_r3\_s04\_h1  	  & 3.0     & $2.5\times10^{7}$   & $1.0\times10^{7}$   & 0.4   & 1.0   & 0.097 & 102.8 & 0.221 & 0.15\\
    m7\_r5\_s04\_h1  	  & 5.0     & $2.5\times10^{7}$   & $1.0\times10^{7}$   & 0.4   & 1.0   & 0.209 & 79.6   & 0.048 & 0.87\\
    m8\_r5\_s04\_h1  	  & 5.0     & $2.5\times10^{8}$   & $1.0\times10^{8}$   & 0.4   & 1.0   & 0.066 & 251.6 & 0.477 & 0.027\\ \hline
    m7\_r3\_s04\_h003& 3.0     & $2.5\times10^{7}$   & $1.0\times10^{7}$   & 0.4   & 0.03 & 0.097 & 102.8 & 0.221 & 0.15\\
    m7\_r5\_s04\_h003& 5.0     & $2.5\times10^{7}$   & $1.0\times10^{7}$   & 0.4   & 0.03 & 0.209 & 79.6   & 0.048 & 0.87\\
    m8\_r5\_s04\_h003& 5.0     & $2.5\times10^{8}$   & $1.0\times10^{8}$   & 0.4   & 0.03 & 0.066 & 251.6 & 0.477 & 0.027\\ \hline
    ONC A			  & 0.345& $1.12\times10^{4}$ & $3.75\times10^{3}$ & 0.33 & $-$  & 0.23    & 6.8     & 0.065 & 9.4\\
    ONC B			  & 0.158& $1.25\times10^{4}$ & $4.17\times10^{3}$ & 0.33 & $-$  & 0.066  & 10.8   & 0.759 & 0.52\\ \hline
  \end{tabular}
\end{table*}

If UCDs indeed are the most massive star clusters, their stellar populations would essentially have formed in a single burst over a time-span of $\approx 1 \, \rm{Myr}$ (cf. \citealt{Elm2000,Har2001}), meaning that their stars evolve almost simultaneously. Considering the high energies involved in massive star evolution, this implies that UCDs with very top-heavy IMFs (with high-mass IMF-slopes $1.0 <\alpha <1.6$, see DKB) could have lost 90\% of their initial stellar mass over a time span of $\approx 40$ Myr (which is the lifetime of the least massive stars that evolve into SNe, cf. the stellar evolutionary grid by \citealt{Sch1992}). If there was residual gas (i.e. gas that was not used up in star formation) in them, which was swept out during this phase of violent star cluster evolution, the mass loss would have been even more pronounced. Such an extensive mass loss shapes the later appearance of a stellar system and may even be critical for its survival, if it happens on a short enough time scale \citep{Boi2003,Fel2005}. However, DKB argued from structural parameters that mass loss on a time scale of $40 \, \rm{Myr}$ for UCDs is probably in the adiabatic regime and therefore inflates them, but does not threaten to dissolve them. It is clear that a numerical study of this issue, including a more detailed treatment of mass loss through stellar evolution and residual gas expulsion, is necessary to confirm these arguments. It is provided in this paper.

\section{Setup}
\label{sec:setup}

\subsection{Initial conditions}
\label{sec:initial}

In the present paper, UCDs are assumed to have formed in the monolithic collapse of a fragmenting gas cloud, and thus in contrast to the model for UCD-formation proposed in \citet{Fel2002} and \citet{Fel2005b}, i.e. the merger of a star cluster complex into a single object (see also \citealt{Kro1998}). This is not to say that the merging of star clusters is completely irrelevant for UCD formation. For instance, the densest part at the centre a proto-UCD could undergo monolithic collapse, while in the outskirts of the proto-UCD a multitude of star clusters is formed, which eventually merge. However, the apparent universality of the IMF in star clusters below the mass-scale of a UCD suggests that a UCD could not have a different IMF, if it is exclusively build up from such systems.

The adopted formation scenario for UCDs thus suggests that they are the most massive star clusters, which implies that their stellar population formed rapidly in $\approx 1 \, \rm{Myr}$ (cf. \citealt{Elm2000,Har2001}). The $\alpha$-enrichment of the UCDs in Virgo reported by \citet{Evs2007} indeed suggests a short time scale for star formation, although for UCDs in other environments, this $\alpha$-enrichment is less pronounced or even absent (cf. fig.~8 in \citealt{Mie2007}). For simplicity, we assume that the stellar populations of UCDs have formed instantaneously instead of over a very short time-span. This is a conservative assumption for the present study, since it focusses on the stability of UCDs. A stellar population that is built up over an extended time-span also releases the total energy it produces (through stellar processes) over a longer time-span. In consequence, the mass loss from UCDs, which is powered by the energy produced by the stellar population, will be slower and therefore less threatening for the stability of the UCD.

\subsubsection{Structural parameters}

The UCD-models are set up with their mass distributed according to the Plummer-model \citep{pl11}. The Plummer-model is the simplest plausible and self-consistent model for a star cluster \citep{Bin1987,Heg2003}. The advantage of using Plummer-models is that all major quantities are analytically accessible.

We choose nine different combinations of initial stellar mass, $M_{*,0}$, initial Plummer-radius, $R_{\rm pl,0}$, star formations efficiencies (SFEs) and heating efficiencies (HEs) for the UCD-models. The choices of the mentioned parameters are detailed below and the considered combinations of them are listed in Table~\ref{tab:ini}, together with some major quantities derived from them.

$M_{*,0}$ is chosen such in the models that a stellar mass of the order of $10^6$ to $10^7 \, \rm{M}_{\odot}$ remains after the evolution of the massive stars has come to an end. This is the range in which the stellar masses of the observed UCDs lie. The chosen values for $R_{\rm pl,0}$ are either 3 pc or 5 pc and thus similar to the observed radii of GCs (eg. \citealt{McL2000} or \citealt{Jor2005}). This leads to initial central densities, $\rho_{\rm pl,0,c}$, ranging from $1.9 \times 10^4 \rm{M}_{\odot} \, \rm{pc}^{-3}$ to $4.8 \times 10^5 \rm{M}_{\odot} \, \rm{pc}^{-3}$ (Table~\ref{tab:ini}). These values for $\rho_{\rm pl,0,c}$ are similar to the ones that have been calculated for Galactic open clusters, such as the Orion Nebula Cluster (ONC), whose initial parameters are discussed in \citet{Kro2001a}. They consider models with $\rho_{\rm pl,0,c}=6.5 \times 10^4 \rm{M}_{\odot} \, \rm{pc}^{-3}$ or $\rho_{\rm pl,0,c}=7.6 \times 10^5 \rm{M}_{\odot} \, \rm{pc}^{-3}$ for that star cluster, as can be calculated from the initial masses and half-light radii given in their table~1. The models discussed here are thus not extreme because of the densities in their central regions, but because of the extension of this central region. This may account for the proposed top-heaviness of the IMF in UCDs, see Section~\ref{sec:encounters}.

Embedded star clusters in the Milky way are thus less extended than the models discussed in this paper (also see \citealt{Lad2003}, their table~1). Note that also GCs were initially less extended than the UCD-models discussed in this paper, unless they lost very little mass since their formation. The reason why smaller $R_{\rm pl,i}$ are not considered here are the extreme initial central densities they would imply for the objects (for instance of the order of $10^7 \, \rm{M}_{\odot} \, \rm{pc}^{-3}$ for $R_{\rm pl,0}=1\, \rm{pc}$; also see Fig.~\ref{fig:Inipar}). Besides, the very small crossing times of such objects would make computations of the evolution very time-consuming while mass loss from them would approach the adiabatic regime, where the behaviour of the cluster can be calculated analytically with equation~(\ref{adiabatic}) below.

The actual values of the SFE and the HE are hard to quantify. In order to get an idea of how these parameters would influence the early evolution of a UCD, vastly different and in some cases extreme values for them are considered in this paper.

The SFE is defined as the fraction of the primordial gas that is converted into stars during a star-forming event within the cluster- or UCD-forming cloud core region. In the UCD-models, it is taken to be 1 or 0.4, the latter value being approximately the upper limit of the SFEs reported for open star clusters \citep{Lad2003}. These high choices for the SFEs in UCDs are motivated by the fact that it would be more difficult to expel the primordial gas from UCDs than from open clusters because of the deep potential wells of UCDs (see also \citealt{Elm1997}). It has even been suggested (e.g. in \citealt{Elm1999} or \citealt{Mur2008}) that all available gas is turned into stars, if the forming stellar system is dense and massive enough. If indeed all star clusters and UCDs form on the same time scale, the star formation rate must be higher in UCDs than in any of the less massive stellar systems. Taking 1 Myr as the characteristic time scale for star formation in these systems, the average star formation rate in UCDs would be 10-100 $\rm{M}_{\odot} \, \rm{yr}^{-1}$ (DKB). Assuming UCDs are essentially star clusters and that star formation is the more rapid the denser the primordial gas is, this could be understood if the primordial gas cloud forming a UCD is, compared to open star clusters, compressed to a higher density during its collapse.

The HE is defined as the fraction of the energy released by stellar processes that actually drives gas out of a star-forming region instead of being radiated away. That is, the HE is the ratio between the kinetic energy of the interstellar medium (ISM) expelled from the stellar system to the total energy inserted into its ISM. The HEs in starbursts have been argued to be near 1 in some studies (e.g. \citealt{Che1985}), while others suggest that only a few percent of the energy inserted into the ISM is turned in kinetic energy of gas leaving the stellar system (e.g. \citealt{Rec2001}; also see \citealt{Mel2004}). For the present paper, HEs of 1  and 0.03 are considered.

A major improvement compared to the rather arbitrary choice of SFEs and HEs made here would clearly be to estimate these parameters in self-consistent modelling of a collapsing gas cloud large enough to form a UCD. This would also clarify how long it would actually take the stellar population to form in such a system, but is currently not a computable option.

\subsubsection{A possible influence of encounters on the IMF}
\label{sec:encounters}

\begin{figure}
\centering
\includegraphics[scale=0.80]{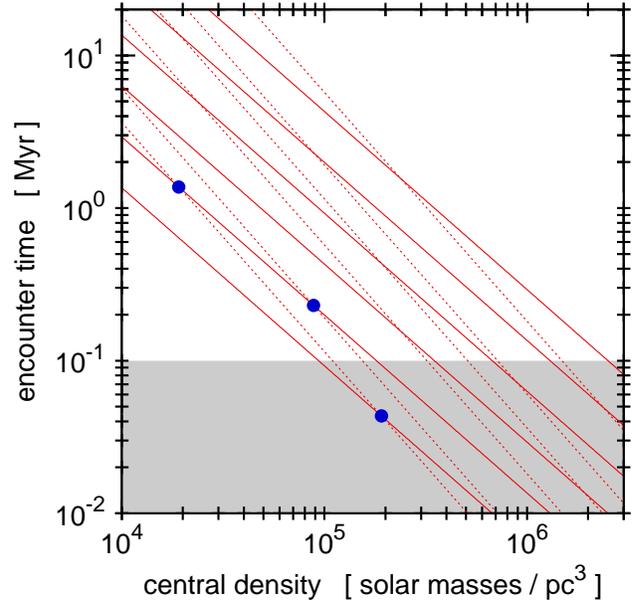}
\caption{The expected time until a proto-star collides with a star, $t_{\rm{enc}}$, in a forming UCD, assuming that half of its total stellar population has already been formed (see Section \ref{sec:encounters} for details). The estimated $t_{\rm{enc}}$ for the UCD-models, in which all gas is eventually converted into stars, are indicated by the three (blue) points. The grey shaded area is where $t_{\rm{enc}}$ is below the approximate life-time of a proto-star, which is assumed to be $10^5$ years. The solid lines show the $t_{\rm{enc}}$ as a function of $\rho_{\rm pl,0,c}$ for different constant $M_{\rm pl,0}$, starting from $10^3 \, \rm{M}_{\odot}$ and increasing by a factor of 10 downwards. The dotted lines show the $t_{\rm{enc}}$ as a function of $\rho_{\rm pl,0,c}$ for different constant $R_{\rm pl,0}$, with $R_{\rm pl,0}$ being 0.1, 0.3, 0.5, 1, 3 and 5 pc from top to bottom.}
\label{fig:Encounters}
\end{figure}

\begin{figure}
\centering
\includegraphics[scale=0.80]{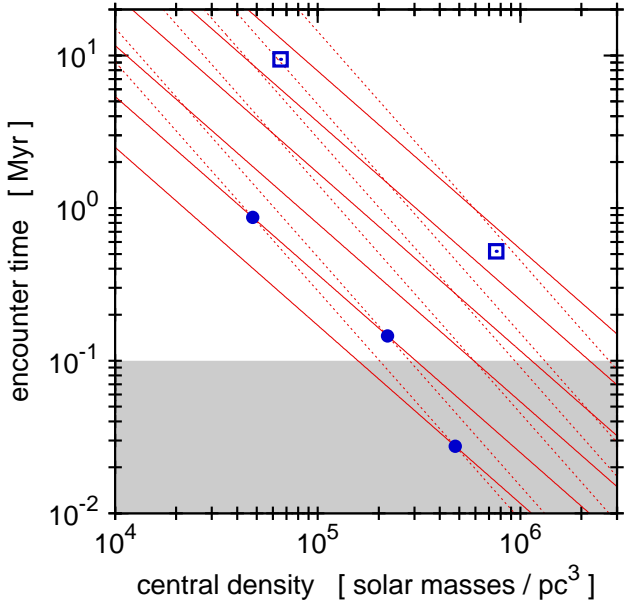}
\caption{As Fig.~(\ref{fig:Encounters}), but for the UCD-models, where a star formation efficiency of 0.4 instead of 1 is assumed. While UCD-models shown here are assumed to have the same \emph{stellar} masses as the ones shown in Fig.~(\ref{fig:Encounters}) their \emph{total} masses  are higher, leading to higher velocity dispersions and thus shorter $t_{\rm{enc}}$ at the same stellar density. The meaning of the solid and the dotted lines is the same as in Fig.~(\ref{fig:Encounters}), but the lowest constant value for $M_{\rm pl,0}$ chosen here is $2.5 \times 10^3 \, \rm{M}_{\odot}$ and increasing by a factor of 10 downwards with every solid line. The open squares show models A and B for the initial states of ONC-type star clusters from \citet{Kro2001a}, which have, compared to the UCD-models in this figure, a slightly lower SFE of 0.33.}
\label{fig:Encounters04}
\end{figure}

If UCDs indeed formed with the initial conditions proposed here, the likeliness for close encounters between members of their emerging stellar populations (stars and proto-stars) would be what sets them apart from ONC-like star clusters. This motivates why the IMF in UCDs might be top-heavy, while this is not observed in star clusters like the ONC.

The case of a proto-star encountering a star is of particular interest. A proto-star exists over a time of $\approx 10^5$ yr until most of its mass has accreted onto the central core \citep{Wuc2003} and is thus short-lived, compared to the characterisic time-scale for star-formation in a star cluster ($\approx 1$ Myr). It has however a radius,  $r_{\rm{proto}}$, of the order of 100 AU for essentially all stellar masses, since the dependency of $r_{\rm{proto}}$ on the mass of the proto-star is only weak (cf. equation~4 in \citealt{Goo2007}). This makes an encounter of a proto-star with a star quite likely, as soon as a considerable stellar population is already present.
 
To estimate a characteristic time-scale for such an encounter for the UCD-models listed in Table~\ref{tab:ini}, consider Plummer-spheres with the initial parameters from that table. Their density-profiles are given as
\begin{equation}
\rho(R)=\frac{3 M_{\rm pl,0}}{4 \pi R_{\rm pl,0}^3} \left[ 1+\left(\frac{R}{R_{\rm pl,0}}\right)^2 \right]^{-\frac{5}{2}}
\label{eq:rhoPlummer}
\end{equation} 
(equation~8.51 in \citealt{Kro2008}), and thus their central \emph{stellar} densities at the time when half of their stellar populations have formed can be estimated as
\begin{equation}
\rho_{\rm pl,*,c}=\frac{3 M_{*,0}}{8 \pi R^3_{\rm pl,0}}.
\label{eq:rhoCentral}
\end{equation}
This density implies a volume that contains one star on average, $V_*$. It can be written as
\begin{equation}
V_*=\frac{\overline{m}}{\rho_{\rm pl,*,c}},
\label{eq:Vstar}
\end{equation}
where $\overline{m}$ is the average stellar mass. The time-dependent volume through which a proto-star in the central region has travelled due to its motion can be written as
\begin{equation}
V(t)=\pi r_{\rm{proto}}^2 \, \sigma_{\rm 3D,0,c} \, t,
\label{eq:Vproto}
\end{equation}
where $\sigma_{\rm 3D,0,c}$ is the \emph{central} 3D velocity dispersion, which is
\begin{equation}
\sigma_{\rm 3D,0,c}=\sqrt{\frac{G M_{\rm pl,0}}{2 R_{\rm pl,0}}},
\label{eq:sigma}
\end{equation}
where $G$ is the gravitational constant (cf. equation~8.59 in \citealt{Kro2008}). The values calculated from equation~(\ref{eq:sigma}) for the UCD-models in this paper are an order of magnitude higher than in the models for the initial states of ONC-type star clusters from \citet{Kro2001a}, while their central densities are essentially the same (see Table~\ref{tab:ini}). The time $t=t_{\rm{enc}}$ by which a proto-star is to be expected to have encountered a star can be calculated be setting $V(t)=V_*$ and solving for $t$. Thus,
\begin{equation}
t_{\rm{enc}}=\frac{\overline{m}}{\pi r_{\rm{proto}}^2 \, \rho_{\rm pl,*,c} \, \sigma_{\rm 3D,0,c}}.
\label{eq:tenc}
\end{equation}
Assuming that a top-heavy IMF results from an canonical IMF by the collisions of proto-stars with stars, $\overline{m}=0.65 \,  \rm{M}_{\odot}$ (which is the average stellar mass for a canonical IMF, see Table~\ref{tab:IMF}), and $r_{\rm{proto}}=100 \, \rm{AU}=4.85\times10^{-4} \, \rm{pc}$ are reasonable choices for emerging open star clusters and emerging UCDs alike. This implies that the IMF would be canonical until it is altered under the influence of encounters between the members of an emerging stellar population and would stay canonical in stellar systems where such encounters are rare at all times.

Note that the derivation of equation~(\ref{eq:tenc}) implicitly assumes that the cross section for an encounter of a proto-star with a star is the geometrical cross section, $A_{\rm{geo}}=\pi r_{\rm{proto}}^2$, whereas the actual cross section for such an encounter is higher due to the influence of gravity. If both the proto-star and the star have the same mass, $\overline{m}$, this actual cross section is given as
\begin{equation}
A=\pi r_{\rm{proto}}^2(1+\Theta)
\label{eq:cross}
\end{equation}
at the centre of the emerging UCD, where $\Theta$ is the Safronov number,
\begin{equation}
\Theta=\frac{2 G \, \overline{m}}{\sigma_{\rm{3D,0,c}}^2 \, r_{\rm{proto}}}
\label{eq:safronov}
\end{equation}
\citep{Mur1996}. However,  assuming $r_{\rm{proto}}=100$ AU and $\overline{m}=0.65 \rm{M}_{\odot}$ leads to $\Theta=0.26$ in the less compact model for the initial states of ONC-type star clusters (ONC A in Table~\ref{tab:ini}). Using the same assumptions, $\Theta$ is lower for all other models in Table~\ref{tab:ini} due to their higher velocity dispersions.  In the case of the UCD-models from this paper, the difference between the actual cross section and the geometric cross section is less than 1 per cent. Thus, gravitational focussing of stars onto the proto-star plays a minor role for the models in Table~\ref{tab:ini}, which justifies the approximation.

The values for $t_{\rm{enc}}$ resulting from equation~(\ref{eq:tenc}) are noted in Table~\ref{tab:ini} and plotted in Figs.~\ref{fig:Encounters} and~\ref{fig:Encounters04}. Comparing these values with the characteristic life-time of a proto-star, $t_{\rm{proto}} \approx 10^5$ years, it can be seen that $t_{\rm{enc}}<t_{\rm{proto}}$ for the UCD-models with $M_{*,0}=10^8 \, \rm{M}_{\odot}$. For the UCD-models with  $M_{*,0}=10^7 \, \rm{M}_{\odot}$ and $R_{\rm pl,0}=3 \, \rm{pc}$, $t_{\rm{enc}}$ is only slightly larger than $t_{\rm{proto}}$. However, for the UCD-models with $M_{*,0}=10^7 \, \rm{M}_{\odot}$ and $R_{\rm pl,0}=5 \, \rm{pc}$,  $t_{\rm{enc}}$ exceeds $t_{\rm{proto}}$ by about an order of magnitude. This suggests that the encounters between proto-stars and stars would influence star-formation in the models with $M_{*,0}=10^8 \, \rm{M}_{\odot}$ and also, to a much lesser extent, in the more compact UCD-models with $M_{*,0}=10^7 \, \rm{M}_{\odot}$, but not in the UCD-models with $M_{*,0}=10^7 \, \rm{M}_{\odot}$ and $R_{\rm pl,0}=5 \, \rm{pc}$. The UCD-models with $M_{*,0}=10^7 \, \rm{M}_{\odot}$ and $R_{\rm pl,0}=5 \, \rm{pc}$ are in this respect similar to models A and B  for the initial states of ONC-type star clusters from \citet{Kro2001a} (see Fig.~\ref{fig:Encounters04}). This implies, invoking the universality of the IMF in open star clusters,  that the IMF in those UCD-models should also be given by equation~(\ref{eq:IMF}), if deviations from the canonical IMF are caused by encounters between proto-stars and stars. According to the calculations in this paper, the UCD-models with $M_{*,0}=10^7 \, \rm{M}_{\odot}$ and $R_{\rm pl,0}=5 \, \rm{pc}$ would indeed only evolve into objects similar to an observed UCD if their mass loss is as implied by the canonical IMF. The IMF in the other UCD-models would however have to be top-heavy to some extent for this (see Section~\ref{sec:results}). The UCD-models in this paper are thus self-consistent in that sense. 

We note that equation~(\ref{eq:tenc}) reveals the particular importance of encounters between proto-stars and stars. For a collision between two stars, $r_{\rm{proto}}$ has to be substituted by a value $\ll 1 \, \rm{AU}$, which leads to $t_{\rm{enc}} \gg 10^5$ years. Thus, collisions between stars only as a mechanism that changes the shape of the IMF \citep{Bon1998,Bon2002} requires even higher densities. For the encounter between two proto-stars, the density of stars at a given time has to be substituted by the density of proto-stars at that time. Taking 1 Myr as the characteristic time-scale on which star-formation takes place and $10^5$ years as the life-time of a proto-star suggests that the density of proto-stars is $\approx 0.1 \rho_{\rm pl,*,c}$, which is five times less than the density of stars at the time when half of the total stellar population of the UCD has formed. 

A caveat to the above discussion is that it is not specified what the consequence of a collision between a proto-star and a star is. This is a merger if the encounter is slow enough. If the encounter is fast enough for the star to only pass through the proto-star, the star transfers some of its kinetic energy on the proto-star and thereby disperses some of the matter that would otherwise accrete on the proto-star. For deciding which of these processes would dominate for a given velocity dispersion, as well as for answering the question of how and to what extent they would alter the IMF, detailed modelling of the collisions would be required. However, the discussion here implies that any process resulting from an encounter between stars and proto-stars should only be relevant for the denser UCD-models in Tab.~\ref{tab:ini}, in contrast to the models for ONC-type star clusters, where most proto-stars should be unaffected by encounters.

We revisit the matter of a possible influence of encounters on the IMF in UCDs in Section~\ref{sec:implications}.

\subsubsection{The IMF of the UCD-models}

\begin{table}
 \centering
 \caption{The IMFs considered for UCDs. The content of columns is the following: Column 1: the identification number of the IMF, Column 2: the slope of the high-mass end of the IMF, $\alpha$ (where 2.3 is the Salpeter slope), Column 3: the upper stellar mass limit, $m_{\rm{max}}$, Column 4: the ratio between the total inital mass of stars more massive than $8 \, \rm{M}_{\odot}$ and the total initial mass of all stars, Column 5: the ratio between the inital number of stars more massive than $8 \, \rm{M}_{\odot}$ and the initial number of all stars, Column 6: the initial mean mass of stars.}
\label{tab:IMF}
\begin{tabular}{llllll} \hline
IMF & $\alpha$ & $m_{\rm{max}} $ & $M_{\rm{hms,0}}/ M_{*,0}$ & $N_{\rm{hms,0}}/ N_{*,0}$ & $\overline{m}$\\
& & [$\rm{M}_{\odot}$] & & & [$\rm{M}_{\odot}$]\\ \hline
1 & 1.1 & 150 & 0.921 & 0.2031 & 10.02 \\
2 & 1.1 & 100 & 0.886 & 0.1830 & 7.16 \\
3 & 1.5 & 100 & 0.719 & 0.0632 & 2.49 \\
4 & 1.9 & 100 & 0.453 & 0.0210 & 1.07 \\
5 & 2.3 & 150 & 0.230 & 0.0072 & 0.65 \\
6 & 2.3 & 100 & 0.213 & 0.0071 & 0.64 \\
\hline
\end{tabular}
\end{table}

For each of the nine sets of models listed in Table~\ref{tab:ini}, six IMFs are considered. They are either canonical or top-heavy to a different degree and have upper mass limits, $m_{\rm{max}}$, of either $100 \, \rm{M}_{\odot}$ or $150 \, \rm{M}_{\odot}$. However, all of them agree with the canonical IMF (equation~\ref{eq:IMF}) for $m<1 \, \rm{M}_{\odot}$. Studies on $m_{\rm{max}}$ suggest that $m_{\rm{max}}=150 \, \rm{M}_{\odot}$ is more realistic than $m_{\rm{max}}=100 \, \rm{M}_{\odot}$ for very massive star clusters and therefore also for UCDs (e.g. \citealt{Mas1998b}, \citealt{Fig1998}, \citealt{Fig2004}). However, the treatment of stellar evolution and its effect on the mass loss from UCDs in this paper (see Section~\ref{sec:massloss}) is based on stellar models that only range up to a $120 \, \rm{M}_{\odot}$ star. Assuming $m_{\rm{max}}=150 \, \rm{M}_{\odot}$ for our models therefore requires extrapolating from the given data, which may be problematic due to the strong dependencies of stellar properties on stellar mass. The emphasis in this paper is therefore on IMFs with $m_{\rm{max}}=100 \, \rm{M}_{\odot}$ (which is also a common choice in simple stellar population models). The impact of the higher $m_{\rm{max}}=150 \, \rm{M}_{\odot}$ is only tested for the canonical IMF and the most top-heavy IMF.

\subsection{Generating the mass loss Tables}
\label{sec:massloss}

\begin{figure}
\centering
\includegraphics[scale=0.80]{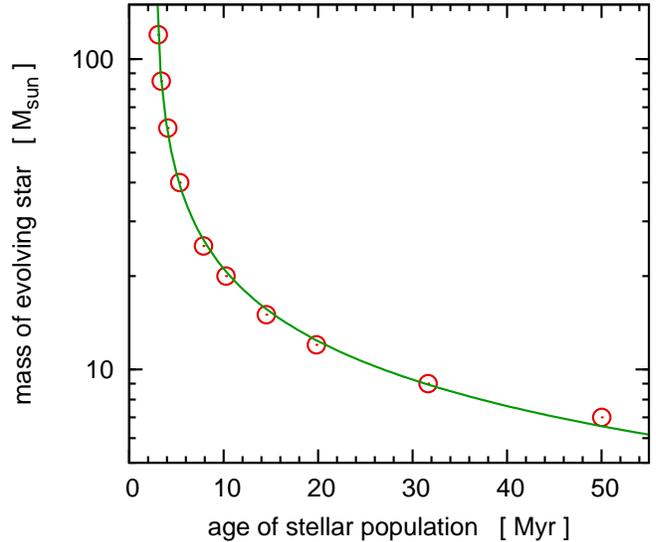}
\caption{The data from \citet{Sch1992} on the lifetimes of stars with different initial masses from 7 to $120 \, \rm{M}_{\odot}$ (open circles) and an interpolation function to them (solid line), which is given by equation~(\ref{eq:lifetimes}). It is apparent that the stars with the highest masses evolve over an extremely short time span. This increases the significance of the upper mass limit of the IMF for the dynamical evolution of a star cluster or UCD.}
\label{fig:Lifetimes}
\end{figure}

\begin{figure}
\centering
\includegraphics[scale=0.80]{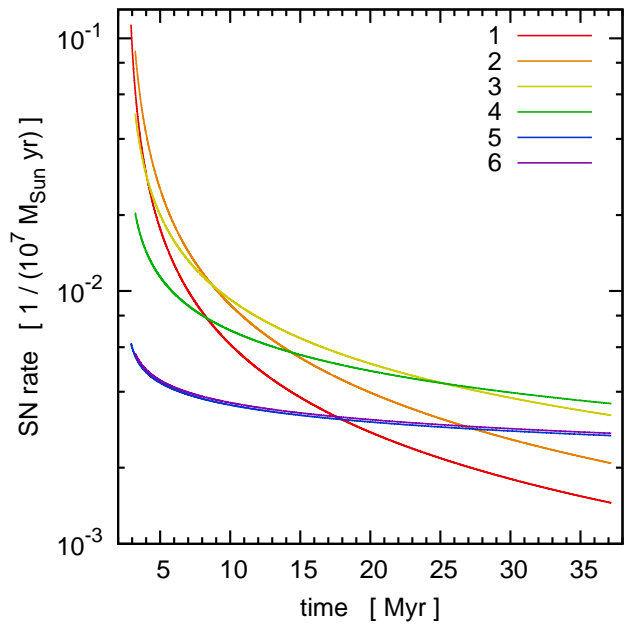}
\caption{ The SN rates with time for the modelled UCDs with initial total stellar mass $M_{*,0}=10^7 \rm{M}_{\odot}$. The different curves are for the different IMFs listed in Tab.~\ref{tab:IMF}. The numbers refer to the labels given to the IMFs in Tab.~\ref{tab:IMF}.  Note that the choice of the upper mass limit of the IMF determines the time when the first SN explodes, but turns out to be almost irrelevant for the SN rates. For the models with $M_{*,0}=10^8 \rm{M}_{\odot}$, the SN rates are higher by a factor of 10. The SN rates are propotional to the energy input by the SN, because all SN are assumed to release the same amount of energy ($10^{51} \rm{erg}$).}
\label{fig:SN}
\end{figure}

\begin{figure}
\centering
\includegraphics[scale=0.80]{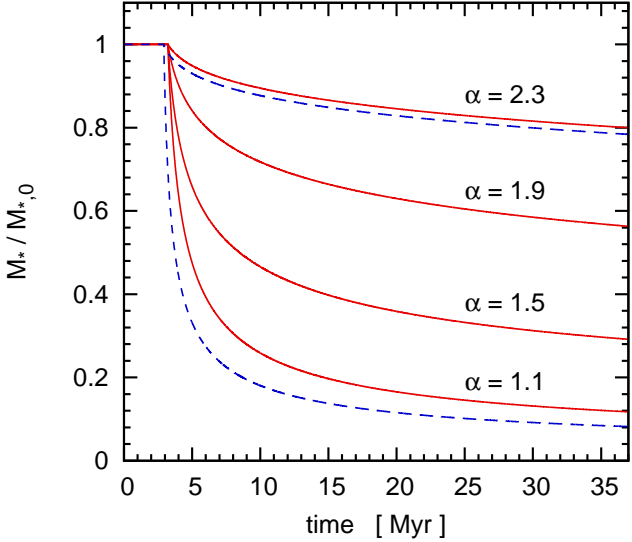}
\caption{The evolution of the stellar mass of the UCD, $M_*$, with time due to stellar evolution. $M_*$ is plotted in units of the initial stellar mass of the UCD, $M_{*,0}$. The different lines in this figure represent $M_*$ for the different IMFs listed in Table~\ref{tab:IMF}. The high-mass slope of the IMF is noted above the curves. Solid (red) lines are for IMFs with an upper mass limit of $100 \, \rm{M}_{\odot}$ and dashed (blue) lines are for an upper mass limit of $150 \, \rm{M}_{\odot}$. The choice of the upper mass limit has only a minor impact for the canonical IMF ($\alpha=2.3$), but is more significant for the most top-heavy IMF we consider ($\alpha=1.1$). For models with SFE=1 and HE=1 these curves show the total mass loss as well.}
\label{fig:Massloss1}
\end{figure}

The interstellar medium (ISM) of a new-born star cluster or UCD is massively heated by the radiation from massive stars, which leads to a mass loss from it until the ISM is depleted. The eventual evolution of the massive stars into supernovae (SNe) heats the ISM as well, but also replenishes the ISM. The rate at which mass is lost from the star cluster or UCD due to this interaction between the massive stars and the ISM is the driving force for its early evolution. This is why the mass loss rate has to be quantified for our models. It is recorded in look-up tables, listing how much the mass of the UCD-models has to be reduced for each time-step in the calculation.

Evidently, knowing the lifetimes of massive stars is essential for generating the mass loss tables. A very good proxy for the time at which the life of the star ends is the time at which carbon burning has finished. The time that has elapsed until this evolutionary stage is reached is taken from the stellar evolutionary grid by \citet{Sch1992} for massive stars with various initial masses. This time-span is identified with the lifetime of a star in this paper. A good fit to the lifetimes of stars with high mass ($m_* \ge 7 \, \rm{M_{\odot}}$) and low metallicity ($Z=0.001$ and $[Z/\rm{H}]=-1.3$ respectively) is the function
\begin{equation}
 m_*=a(t_*-b)^c,
\label{eq:lifetimes}
\end{equation}
with
\begin{eqnarray}
 \nonumber a=74.6,\quad
 \nonumber b=2.59,\quad
 \nonumber c=-0.63.
\end{eqnarray}
where the initial mass of the star, $m_*$, is measured in $\rm{M}_{\odot}$ and the lifetime of the star, $t_*$, is measured in Myr (Fig.~\ref{fig:Lifetimes}). It thus covers the whole range of stellar initial masses of stars that undergo SN explosions at the end of their evolution, which is $m_* \apprge 8 \, \rm{M_{\odot}}$ \citep{Koe1996}. The parameters $a$, $b$ and $c$ have been found by a least-squares fit. The models for low-metallicity stars were preferred over models for stars with Solar metallicity because of the mostly sub-solar metallicities of the UCDs \citep{Mie2006,Evs2007}. This choice has however only a minor impact on the best-fitting parameters $a$, $b$ and $c$. Note that since all stars in the UCD-models are assumed to have formed at once in this paper, it is possible to substitute the stellar lifetime, $t_*$, in equation~(\ref{eq:lifetimes}) with the age of the UCD-model, $t$, in order to find the initial mass of the stars that undergo SNe at that time.

Now consider the increase of the age of the UCD-model by the time step $t_{\rm{i}} \rightarrow t_{\rm{i+1}}$. During this time, $\Delta N_*$ stars with a total mass $\Delta M_*$ will complete their evolution. These quantities can be written as
\begin{equation}
\Delta N_{\rm{*,i}}= \frac{M_{*,0}}{\rm{M}_{\odot}} \int_{m_{*\rm{,i+1}}}^{m_{*\rm{,i}}} \xi(m_*) \, dm_*,
\label{eq:dN}
\end{equation}
and
\begin{equation}
\Delta M_{\rm{*,i}}= \frac{M_{*,0}}{\rm{M}_{\odot}} \int_{m_{*\rm{,i+1}}}^{m_{*\rm{,i}}} \xi(m_*) m_* \, dm_*,
\label{eq:dM}
\end{equation}
where $\xi (m_*)$ is the IMF, $m_{*\rm{,i}}$ is the initial mass of stars that evolve at $t=t_{\rm{i}}$ and $m_{*\rm{,i+1}}$ is the initial mass of stars that evolve at $t=t_{\rm{i+1}}$. $M_{*,0}$ is the total initial stellar mass of the UCD-model. Given the normalisation chosen for the IMF (see equations~\ref{eq:IMF} and~\ref{eq:norm1}), the purpose of the factors $M_{*,0}/\rm{M}_{\odot}$ is to scale equations~\ref{eq:dN},~\ref{eq:dM} and~\ref{eq:lum} to a UCD-model with the initial mass of $M_{*,0}$. $\Delta N_*$ is equivalent to the number of SNe during the time step $t_{\rm{i}} \rightarrow t_{\rm{i+1}}$; i.e. the SN-rate in the limit of $t_{\rm{i+1}}-t_{\rm{i}} \rightarrow 0$. At the time when the most massive stars evolve, this SN-rate is, for instance, $\approx 1$ SN per 10 years for the UCD-models with $M_{*,0}=10^7 \, \rm{M}_{\odot}$ and a high-mass IMF slope of $\alpha=1.1$, while it is a few SN per $10^3$ years for the UCD-models with $M_{*,0}=10^7 \, \rm{M}_{\odot}$ and $\alpha=2.3$. The influence of the top-heavyness of the IMF on the SN-rates decreases as time proceeds. The SN-rates for the UCD-models with $M_{*,0}=10^7 \, \rm{M}_{\odot}$ and the IMFs from Table~\ref{tab:IMF} are shown in Fig.~\ref{fig:SN}. The SN-rates for the UCD-models with $M_{*,0}=10^8 \, \rm{M}_{\odot}$ are higher by a factor of 10 compared to the ones shown in this figure, but the same otherwise.

Fig.~\ref{fig:Massloss1} depicts the change of the stellar mass of the UCD-model with time, i.e. $M_{\rm{*,i}}=M_{*,0}$ for $t_{\rm{i}} \le t_{\rm{SN}}$ and $M_{\rm{*,i}}=M_{*,0}-\sum _{n=1}^i \Delta M_{\rm{*,n}}$ for $t_{\rm{i}}>t_{\rm{SN}}$, where $M_{*,0}$ is the total initial stellar mass and $t_{\rm{SN}}$ is the time when the first stars become SNe.

The total energy deposited by stars into their surroundings by radiation and stellar winds at the time $t=t_i$, $L_{\rm{*,i}}$, is given as
\begin{equation}
 L_{\rm{*,i}}=\frac{M_{*,0}}{\rm{M}_{\odot}} \int_{0.1}^{m_{\rm{max,i}}} \xi(m_*)l(m_*) \, dm_*,
\label{eq:lum}
\end{equation} with $m_{\rm{max,i}}$ being the mass of the most massive star that has not evolved into a SN at that time. $l(m_*)$ is the energy deposition rate of stars into the ISM through radiation and stellar winds as a function of their initial mass. It is estimated as
\begin{equation}
 l(m_*)=2.16\times 10^{47}\left( \frac{m_*}{\rm{M}_{\odot}}\right)^{1.72} \frac{\rm{erg}}{\rm{Myr}},
 \label{eq:masslum}
\end{equation}
which is identical to equation~(12) in \citet{Bau2008}. As in \citet{Bau2008}, equation~(\ref{eq:masslum}) is applied to stars of all masses, even though it was obtained in a fit to high-mass stars. Note that the positive exponent in equation~(\ref{eq:masslum}) and the negative exponent in the IMF cancel out more or less in equation~(\ref{eq:lum}). The contribution of low- and intermediate-mass stars to the total energy deposition into the ISM is therefore small at first, because the masses of high-mass stars are distributed over a much wider range.

\begin{figure*}
\centering
\includegraphics[scale=0.80]{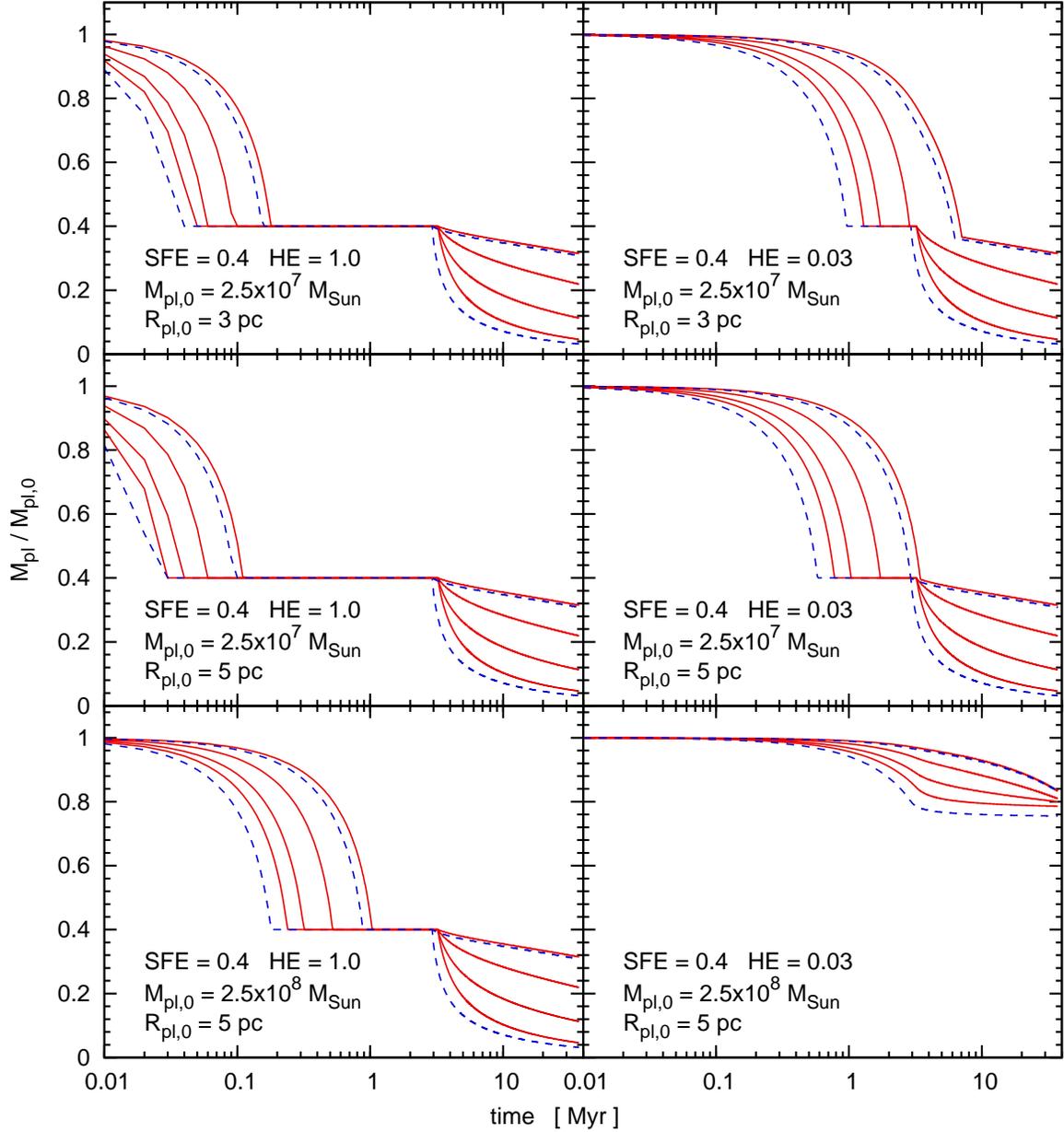}
\caption{The change of the total mass of the UCD-model, $M_{\rm{pl}}$, with time due to stellar radiation and evolution for our models with $\rm{SFE}=0.4$ and $\rm{HE}=1.0$ or $\rm{HE}=0.03$ relative to the total initial mass of the UCD-model, $M_{\rm{pl,0}}$. The assumptions regarding SFE, HE, initial mass of the UCD-model and inital Plummer-radius of the UCD-model for the mass loss histories shown are indicated in the corresponding panel. The different lines in this figure represent $M_{\rm{pl}}$ for the different IMFs listed in Table~\ref{tab:IMF}. The mass loss increases with the top-heavyness of the assumed IMF: for the topmost curves $\alpha=2.3$ and for the lowermost curves $\alpha=1.1$. Solid (red) lines are for IMFs with an upper mass limit of $100 \, \rm{M}_{\odot}$ and dashed (blue) lines are for an upper mass limit of $150 \, \rm{M}_{\odot}$. The choice of the upper mass limit has only a minor impact for the canonical IMF ($\alpha=2.3$), but is much more significant for the most top-heavy IMF that are considered ($\alpha=1.1$). Note that for $t \apprge 3\, \rm{Myr}$ the mass loss histories shown in this figure are equal to the change in stellar mass shown in Fig.~\ref{fig:Massloss1} if the primordial gas is expelled before the first star has evolved completely. However, contrary to Fig.~\ref{fig:Massloss1} the time-axis is scaled logarithmically here, in order to show the sometimes very rapid expulsion of the primordial gas.}
\label{fig:Massloss2}
\end{figure*}

\subsubsection{The algorithm}
\label{sec:algorithm}

The integrations in equations~(\ref{eq:dN}) to~(\ref{eq:lum}) are done numerically for the IMFs listed in Table~(\ref{tab:IMF}). The used program is structured as outlined below.

Start at $t=0$ with a set of initial parameters taken from Table~\ref{tab:ini} and an IMF taken from Table~\ref{tab:IMF}. Let $\Delta t$ be the time step from $t_{\rm{i}}$ to $t_{\rm{i+1}}$ and $t_{\rm{SN}}$ the time when the first stars become SNe.
\begin{enumerate}
 \item If $t_{\rm{i}}>t_{\rm{SN}}$, calculate which stars evolve from $t=t_{\rm{i}}$ to $t=t_{\rm{i}}+\Delta t$  using equation~(\ref{eq:lifetimes}) and then which total mass these stars have, $\Delta M_{*,\rm{i}}$ (equation~\ref{eq:dM}). This mass is added to the total mass of the interstellar medium, $M_{\rm{ISM,i}}$. This includes the possibility that the UCD-model had no ISM left at the end of the previous time step. In this case, $M_{\rm{ISM,i}}=\Delta M_{*,\rm{i}}$.
 \item The rate at which the stars and the SNe deposit energy into the ISM during the time step is calculated, $L_{\rm{i}}$. $L_{\rm{i}}$ is approximated by $L_{\rm{i}}=L_{\rm{*,i+1}}+L_{\rm{SN,i}}$, where $L_{\rm{*,i+1}}$ is the rate at which the stars deposit energy into the ISM at $t=t_{\rm{i+1}}$ (equation~\ref{eq:lum}) and $L_{\rm{SN,i}}$ is the energy that the SNe deposit into the ISM from $t=t_{\rm{i}}$ to $t=t_{\rm{i}}+\Delta t$ (which is in the limit of $\Delta t \rightarrow 0$ an energy deposition \emph{rate} as well). Using the number of stars that evolve during the time step $\Delta N_{\rm{*,i}}$, (equation~\ref{eq:dN}), $L_{\rm{SN,i}}$ can be estimated by assuming that each SN releases a characteristic amount of kinetic and electromagnetic energy, which are the forms of energy that are relevant for driving matter out of the UCD-model. Estimating this quantity as $10^{51} \, \rm{erg}$ per SN (e.g. \citealt{Car1996}) leads to $L_{\rm{SN,i}}=10^{51}\Delta N_{\rm{*,i}}\, \rm{erg} \, \rm{Myr}^{-1}$. The total luminosity is multiplied by the HE to obtain $L_{\rm{kin,i}}$, which is the luminosity that is not radiated away through thermal emission of the ISM, but is converted into kinetic energy of the gas leaving the UCD.
 \item The time $\tau_{\rm{i}}$ it would take until all gas is expelled from the UCD-model is estimated, assuming that  $L_{\rm{kin,i}}$ does not change during that time. This is done using the equation $\tau_{\rm{i}} L_{\rm{kin,i}}=|E_{\rm{pot,i}}-E_{\rm{pot*,i}}|$, where $E_{\rm{pot,i}}$ is the total binding energy of the UCD-model at $t=t_i$ and $E_{\rm{pot*,i}}$ is the binding energy the UCD would have if it would lose all gas at that time. Note that the UCD-model inflates as it loses mass and $E_{\rm{pot*,i}}$ should therefore be calculated using the Plummer-radius the UCD-model has after all gas is expelled. We estimate it using the relation between initial radius and final radius of a stellar system for adiabatic mass loss, even though the mass loss is in our case sometimes clearly not adiabatic. This relation is given by
\begin{equation}
\frac{r_{\rm{final}}}{r_{\rm{init}}}=\frac{M_{\rm{init}}}{M_{\rm{final}}},
\label{adiabatic}
\end{equation}
where $r_{\rm{init}}$ and $M_{\rm{init}}$ are radius and mass of the stellar system at the beginning of mass loss, respectively, and $r_{\rm{final}}$ and $M_{\rm{final}}$ the are radius and mass of the stellar system at the end of mass loss, respectively (\citealt{Kro2008} and references therein). Equation~(\ref{adiabatic}) underestimates the expansion of the UCD-models at times of non-adiabatic mass loss (cf. equation~8.20 in \citealt{Kro2008}). Therefore, if the UCDs have experienced extended non-adiabatic mass loss, the mass loss rates calculated here are too low, because a more pronounced expansion implies that the potential well becomes shallower and the remaining gas requires less energy to escape from it. This is even more important for a more top-heavy IMF.
 \item If $\tau_{\rm{i}}<\Delta t$, set $\tau_{\rm{i}}=\Delta t$. The mass loss of the UCD-model during the time step is assumed to be
 \begin{equation}
 \delta M=M_{\rm{ISM,i}}\frac{\Delta t}{\tau_{\rm{i}}}.
 \end{equation}
 \item Calculate the new parameters of the UCD-model after it has lost the mass $\delta M$: The new total stellar mass is decreased by $\Delta M_{*\rm{i}}$ and the new Plummer-radius is estimated using equation~(\ref{adiabatic}).
 \item If $t_{\rm{i+1}}=t_{\rm{i}}+\Delta t$ is less than it takes a star with $m_*=8 \, \rm{M}_{\odot}$ to evolve into a SN according to equation~(\ref{eq:lifetimes}), repeat steps (i) through (vi), but for $t_{\rm{i+1}}$ instead of $t_{\rm{i}}$.
\end{enumerate}

The underlying assumption in the chosen approach is that the material expelled from a SN does not \emph{immediately} escape the UCD, but that its kinetic energy is thermalised, as it is assumed for massive star clusters in, e.g., \citet{Ten2007}. This can happen either through interaction with the surrounding ISM or through the collision of the expanding envelopes of different SNe. The latter becomes more relevant with increasing top-heaviness of the IMF. The notion of the thermalisation of the SN ejecta is flawed if there is no ISM left and if the SNe are too few for their envelopes to interact with one another. However, in this case also very low HEs are sufficient to keep the UCD-models gas-free.

The mass loss histories calculated by using the above routine are shown in Fig.~(\ref{fig:Massloss2}) for the models with SFE=0.4 and HE=1 or HE=0.03. For the the models with SFE=1 and HE=1, the evolution of the stellar mass of the UCD-models shown in Fig.~\ref{fig:Massloss1} also illustrates their mass loss history, since the UCD-models are gas-free at all times in this case.

\subsubsection{The role of compact stellar remnants}
\label{sec:remnants}

\begin{table}
 \centering
 \caption{The total masses of all stars more massive than $8 \, \rm{M}_{\odot}$ ($M_{\rm{hms,0}}$, Column~2) and the total masses of their compact remnants ($M_{\rm{rem}}$, Columns~3 and~4) for the IMFs in Table~\ref{tab:IMF}. The masses are in units of the total mass of all stars that were formed initially, $M_{*,0}$. The two different values for $M_{\rm{rem}}$ for a given IMF reflect that the mass of the remnants of very massive stars is poorly known. While the mass of the compact remnants of stars with initial masses of $8 \, \rm{M}_{\odot} \le m_* < 25 \, \rm{M}_{\odot}$ is $1.35 \, \rm{M}_{\odot}$ in both estimates, the mass of the compact remnants of stars with $m_*>25 \, \rm{M}_{\odot}$ is assumed to be either $0.1m_*$ (Column~3) or $0.5m_*$ (Column~4).}
\label{tab:remnants}
\begin{tabular}{llll} \hline
IMF & $M_{\rm{hms,0}}/ M_{*,0}$ & $M_{\rm{rem}}/ M_{*,0}$ & $M_{\rm{rem}}/ M_{*,0}$ \\
& &$m_{\rm{BH}}=0.1 m_*$ & $m_{\rm{BH}}=0.5 m_*$\\ \hline
1 & 0.921 & 0.0910 & 0.409\\
2 & 0.886 & 0.0871 & 0.369\\
3 & 0.719 & 0.0709 & 0.271\\
4 & 0.453 & 0.0452 & 0.150\\
5 & 0.230 & 0.0234 & 0.0698\\
6 & 0.213 & 0.0218 & 0.0605\\
\hline
\end{tabular}
\end{table}

A simplification that is made in the creation of the mass loss histories is that the whole mass of the evolved stars is added to the ISM, including the mass of their compact remnants. For testing under which conditions this approximation is reasonable, the total mass of all compact remnants of stars with initial masses $m_*>8 \, \rm{M}_{\odot}$, $M_{\rm{rem}}$, needs to be compared to the total mass of their progenitors, $M_{\rm{hms,0}}$. If an IMF is given, calculating  $M_{\rm{rem}}$ requires a relation between the initial masses of stars and the masses of their compact remnants. Such an initial-to-final mass relation is, e.g., formulated in equation~(8) of DKB. Their equation is also used here. Thus, stars with initial masses of $8 \, \rm{M}_{\odot} \le m_* < 25 \, \rm{M}_{\odot}$ are thought to evolve into neutron stars (NSs) with a mass of $1.35 \, \rm{M}_{\odot}$, which is the mass \citet{Tho1999} have found for pulsars, i.e. a sample of neutron stars that can easily be detected. Stars with even higher initial masses are believed to evolve into black holes, but the actual masses of these black holes (BHs) are poorly constrained (cf. figs.~12 and~16 in \citealt{Woo2002}). Therefore, two cases are considered for the masses of BHs, namely the case that they all have 10 per cent of the initial mass of their progenitors ($m_{\rm{BH}}=0.1 m_*$) and the case that they have 50 per cent of the initial mass of their progenitors ($m_{\rm{BH}}=0.5 m_*$) are considered. The resulting values are noted in Table~\ref{tab:remnants}.

It is apparent from these numbers that for $m_{\rm{BH}}=0.1 m_*$ the mass locked up in compact remnants is indeed negligible, while this is not the case for $m_{\rm{BH}}=0.5 m_*$. However, the masses of observationally confirmed stellar-mass BHs (see \citealt{Cas2007}) seem to favour the case of  $m_{\rm{BH}}=0.1 m_*$, leading to BH masses $\apprle 10 \, \rm{M}_{\odot}$. Apart from that, \citet{Lyn1994} report a mean birth velocity of $450\pm90 \, \rm{km \, s^{-1}}$ for pulsars (i.e. neutron stars) and the processes that precede the birth of a stellar mass BH are essentially the same as the ones that precede the birth of a neutron star \citep{Woo2002}. This suggests that a large fraction of the compact remnants (BHs as well as neutron stars) are born with velocities well above the escape velocity of 
the UCD-models in Tab.1, which for a Plummer sphere is about twice the velocity dispersion (compare equations~8.59 and~8.61 in \citealt{Kro2008}). Thus, in a realistic scenario, the total mass of the compact remnants \emph{remaining} in the UCD is likely to be small compared to the total mass of the progenitors of \emph{all} compact remnants.  Moreover, the mass-loss histories created by the algorithm described in Section~\ref{sec:algorithm} suggest that the UCDs are gas-free at the end of the evolution of massive stars, with the exception of the models with high initial mass and low heating efficiency. The latter models suggest however that the UCDs consist mainly of gas at that time, which seems unlikely (see Section~\ref{sec:h003s04}). As a conclusion, UCDs are likely to have lost most of the mass that was locked in massive stars at the time when massive stars have evolved,  if they formed as is assumed here (i.e. as very massive star clusters). This mass loss proceeds however not only by the escape of the gaseous components of the SN-remnants from the UCD (i.e. a process modelled by the algorithm in Section~\ref{sec:algorithm}), but also by the ejection of the compact remnants. This latter process would play a substantial role for the total mass loss of the UCD if $m_{\rm{BH}}=0.5 m_*$, but not if $m_{\rm{BH}}=0.1 m_*$.
 
Note that the expectation of a large difference between the total mass of the compact remnants left in UCDs and the total mass of their progenitors is also the reason why the IMFs of the UCDs have to be so extremely top-heavy, if the enhanced $M/L_V$ ratios of the UCDs are to be explained by an over-abundance of stellar remnants. This is what motivated the sometimes extreme choices for the IMF in the UCD-models in the first place.

In essence, neglecting the remnant masses seems justifiable in the context of the present study, as it also helps to avoid a number of very speculative assumptions. This includes the precise mass of the remnants, which fraction of them remains in the UCDs and how much of the kinetic energy available from the SNe is tranferred to them. As a result, the mass loss is over-estimated in the UCD-models, but probably not by much more than 10 per cent. Consequently, their expansion is over-estimated as well by about the same amount, if the heating is sufficient to expel all gas from them. This bias is thus opposed to the bias induced by the assumption of adiabatic mass loss at all times in the calculation of the mass loss histories.

We note that also the treatment of the energy input from SNe (each of them contributes $10^{51} \, \rm{erg}$) and the energy input from stars (equation~\ref{eq:lum} using equation~\ref{eq:masslum}) is only approximate, but can hardly be done with greater precision with current knowledge.

\subsection{Time evolution of the UCDs}
\label{sec:evolution}

The UCD-models are set up to be in virial equilibrium before the onset of mass loss. This is motivated by the fact that star-formation in a star cluster takes place on a time-scale of $\approx1$ Myr, while the crossing times in the UCD-models are about an order of magnitude lower (see Table~\ref{tab:ini}) and the time-scale for violent relaxation is a few crossing times \citep{Bin1987}. Thus, the time-scale for the UCD-model to settle into a state near virial equilibrium is shorter than the time-scale for the formation of its stellar population. Note that the assumption of virial equilibrium is crucial for the validity of the results in this paper, since UCDs would evolve completely different if they were not in virial equilibrium at the onset of gas expulsion, see \citet{Goo2009}. But it is also argued there that very massive star clusters are much more likely to be in virial equilibrium at that time.

To calculate the evolution of the UCDs in the first few Myr the particle-mesh code Superbox \citep{fe00} is used. Each UCD is represented by 1 million particles and is integrated forward in time until $200$~Myr using a small time-step of $0.01$~Myr for the models with HE=1 and SFE=1 and a time-step of $0.005$~Myr for all models with SFE=0.4. The smaller time-step for the models with SFE=0.4 is necessary because of their shorter crossing times due to their higher initial masses for our assumed stellar masses, see Table~\ref{tab:ini}. The code is altered to allow for the mass loss due to gas expulsion and rapid stellar evolution in the first tens of Myr. To mimick this mass loss we implemented the look-up tables whose generation is described in Section~\ref{sec:massloss}. They give the total mass of the UCD at each time-step. The mass of each particle and henceforth the total mass of the modelled UCD is reduced accordingly.

The UCDs are modelled in isolation, i.e. in the absence of a tidal field, even though UCDs are found in the vicinity of massive elliptical galaxies. But regarding the short time span of our computations of 200 Myr (compared to their orbital times of a Gyr or longer) and the fact that e.g.\ at a distance of $80$~kpc and adopting the potential of M\,82 the tidal radii would be $600$~pc for the models with $M_{\rm{pl,0}}=10^7 \, \rm{M}_{\odot}$ and $1400$~pc for the models with $M_{\rm{pl,0}}=10^8 \, \rm{M}_{\odot}$, the effect of the tidal fields can be neglected. (See also table~6 in \citealt{Hil2007} and table~8 in \citealt{Evs2007} for a comparison between half-light radii and tidal radii of UCDs.)

\section{Results}
\label{sec:results}

We calculated a suite of 56 models, combining each of the sets of UCD parameters given in Table~\ref{tab:ini} with each IMF given in Table~\ref{tab:IMF}. The results are discussed separately for the different assumptions regarding the SFE and the HE in the following, see also Tables~3,~4 and 5.

\subsection{SFE=1}
\label{sec:h1s1}

\begin{figure}
  \centering
  \epsfxsize=8cm
  \epsfysize=8cm
  \epsffile{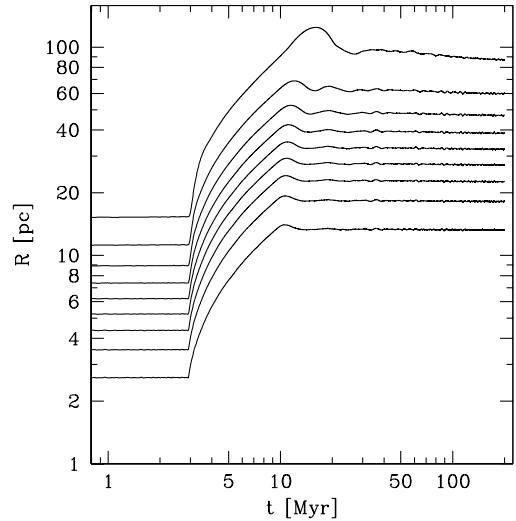}
  \caption{Change of the Lagrangian radii (10, 20, ... 90~\% mass) with time for model m8\_r5\_s1\_h1 with IMF 1 ($m_{\rm{max}}=150 \, \rm{M}_{\odot}$, $\alpha=1.1$).}
  \label{fig:lagrad}
\end{figure}

\begin{figure}
\centering
\includegraphics[scale=0.80]{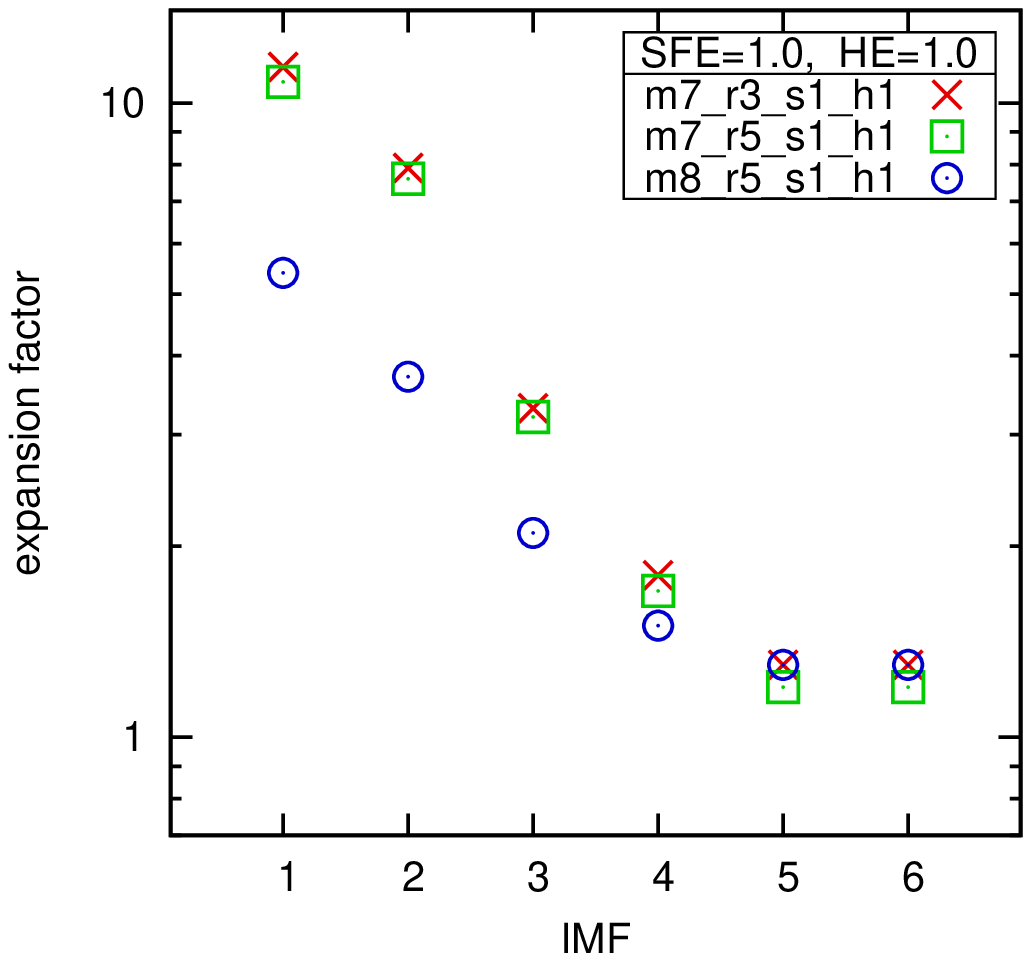}
\caption{Expansion factors $f_{\rm e}$ of our models for a star formation efficiency of 1 and a heating efficiency of 1 plotted against the number given to the assumed IMF (as in Table~\ref{tab:IMF}). The symbols show the different initial structural parameters of the UCD-models: (red) crosses for $R_{\rm pl,0}=3 \, \rm{pc}$ and $M_{\rm pl,0}=10^7 \, \rm{M}_{\odot}$, (green) squares for $R_{\rm pl,0}=5 \, \rm{pc}$ and $M_{\rm pl,0}=10^7 \, \rm{M}_{\odot}$ and (blue) circles for $R_{\rm pl,0}=5 \, \rm{pc}$ and $M_{\rm pl,0}=10^7 \, \rm{M}_{\odot}$.}
\label{fig:exph1s1}
\end{figure}

\begin{table*}
  \centering
  \caption{Final results of the calculations for a SFE of 1 and a HE of 1. The models whose final parameters match the observed parameters of UCDs best are marked with a (+) before the first column. The information given in the columns is the following: Column~1: the name of the model as given in Table~\ref{tab:ini}, Column~2: the IMF as given in Table~\ref{tab:IMF}, Column~3: the total mass of stars that have not evolved at the end of massive-star evolution (i.e. stars with $m<8 \, \rm{M}_{\odot}$) in units of the initial total mass of all stars, Column~4: the total mass of stars that remain bound to the cluster in units of the total mass of all stars less massive than $8 \, \rm{M}_{\odot}$ (i.e. no stars become unbound if the entry in this column is 1), Column~5: the mass of the cluster at the end of the calculation, Column~6: the final half-mass radius, Columns~7 and~8: the final Plummer-radius $R_{\rm pl,f}$ and its $1$-$\sigma$ error, Column~9: the expansion factor $f_{\rm e}$, Columns~10 and~11: the central surface density $\Sigma_{0,\rm{f}}$ with its error, Columns~12 and~13: and the central line-of-sight velocity dispersion $\sigma_{0,\rm{f}}$ with its error. $\Sigma_{0,\rm{f}}$ and $\sigma_{0,\rm{f}}$ are derived by fitting Plummer-profiles to the data at $t = 200$~Myr, using a non-linear least-squares Marquardt-Levenberg algorithm. Both the fits to $\Sigma_{0,\rm{f}}$ and to $\sigma_{0,\rm{f}}$ also deliver estimates for $R_{\rm pl,f}$. The quoted value for $R_{\rm pl,f}$ is the one obtained from the fit to $\Sigma_{0,\rm{f}}$, but the one obtained from the fit to $\sigma_{0,\rm{f}}$ is not much different.}
  \label{tab:h1s1}
  \begin{tabular}{lllllllllllllll} \hline
    & model & IMF & $M_{*,\rm{f}}/M_{*,0}$ & $M_{*\rm{b,f}}/M_{*,f}$ & $M_{\rm{f}}$ & $R_{50\rm{,f}}$ & $R_{\rm pl,f}$ & error & $f_{\rm e}$ & $\Sigma_{0,\rm{f}}$ & error & $\sigma_{0,\rm{f}}$ & error \\
    & & & & & [$10^6 \rm{M}_{\odot}$] & [pc] & \multicolumn{2}{c}{[pc]} & & \multicolumn{2}{c}{[M$_{\odot}$\,pc$^{-2}$]} & \multicolumn{2}{c}{[km\,s$^{-1}$]} \\ \hline
    & m7\_r3\_s1\_h1 & 1 & 0.079 & 0.952 & 0.75 & 30.6 & 34.1 & 0.2 & 11.4 & 241 & 1 & 4.19 & 0.02 \\
    & m7\_r5\_s1\_h1 & 1 & 0.079 & 0.850 & 0.67 & 52.2 & 53.9 & 0.9 & 10.8 & 17 & 0 & 3.02 & 0.02 \\
    & m8\_r5\_s1\_h1 & 1 & 0.079 & 0.998 & 7.8 & 26.2 & 27.1 & 0.1 & 5.4 & 3869 & 2 & 21.65 & 0.07 \\
    \hline
    & m7\_r3\_s1\_h1 & 2 & 0.114 & 0.973 & 1.11 & 22.3 & 23.7 & 0.1 & 7.9 & 736 & 2 & 6.09 & 0.02 \\
    & m7\_r5\_s1\_h1 & 2 & 0.114 & 0.953 & 1.09 & 37.4 & 38.0 & 0.2 & 7.6 & 280 & 1 & 4.77 & 0.02 \\
    (+) & m8\_r5\_s1\_h1 & 2 & 0.114 & 1.000 & 11.4 & 18.4 & 18.6 & 0.1 & 3.7 & 12069 & 22 & 31.36 & 0.07 \\
    \hline
    & m7\_r3\_s1\_h1 & 3 & 0.281 & 0.999 & 2.81 & 9.6 & 10.0 & 0.0 & 3.3 & 10324 & 11 & 14.52 & 0.06 \\
    & m7\_r5\_s1\_h1 & 3 & 0.281 & 0.998 & 2.80 & 16.1 & 16.0 & 0.1 & 3.2 & 4131 & 8 & 11.60 & 0.04 \\
    (+) & m8\_r5\_s1\_h1 & 3 & 0.281 & 1.000 & 28.1 & 10.8 & 10.7 & 0.0 & 2.1 & 90300 & 120 & 55.20 & 0.30 \\
    \hline
    (+) & m7\_r3\_s1\_h1 & 4 & 0.547 & 1.000 & 5.47 & 5.1 & 5.3 & 0.0 & 1.8 & 70370 & 170 & 27.50 & 0.02 \\
    (+) & m7\_r5\_s1\_h1 & 4 & 0.547 & 1.000 & 5.47 & 8.6 & 8.5 & 0.0 & 1.7 & 28286 & 27 & 21.90 & 0.30 \\
    & m8\_r5\_s1\_h1 & 4 & 0.547 & 1.000 & 54.7 & 7.4 & 7.4 & 0.0 & 1.5 & 363990 & 310 & 82.50 & 0.80 \\
    \hline
    & m7\_r3\_s1\_h1 & 5 & 0.770 & 1.000 & 7.70 & 3.8 & 3.9 & 0.0 & 1.3 & 181300 & 800 & 37.60 & 0.50 \\
    (+) & m7\_r5\_s1\_h1 & 5 & 0.770 & 1.000 & 7.70 & 6.3 & 6.2 & 0.0 & 1.2 & 73630 & 140 & 30.20 & 0.30 \\
    & m8\_r5\_s1\_h1 & 5 & 0.770 & 1.000 & 77.0  & 6.3 & 6.4 & 0.0 & 1.3 & 678800 & 1400 & 98.30 & 1.10 \\
    \hline
    & m7\_r3\_s1\_h1 & 6 & 0.787 & 1.000 & 7.87 & 3.7 & 3.8 & 0.0 & 1.3 & 191560 & 960 & 38.40 & 0.20 \\
    (+) & m7\_r5\_s1\_h1 & 6 & 0.787 & 1.000 & 7.87 & 6.2 & 6.1 & 0.0 & 1.2 & 78050 & 440 & 30.70 & 0.30 \\
    & m8\_r5\_s1\_h1 & 6 & 0.787 & 1.000 & 78.7 & 6.3 & 6.3 & 0.0 & 1.3 & 713800 & 1100 & 100.30 & 1.20 \\
    \hline
  \end{tabular}
\end{table*}

The assumptions SFE=1 and HE=1 stand for the case of highly efficient star-formation and heating. There is no expulsion of primordial gas in this case, but the mass loss through the evolution of massive stars can still be quite severe (as the UCD is cleared easily from the products of stellar evolution with such a high HE). It amounts to up to about $90$~per cent of the initial mass for the UCDs with the most top-heavy IMFs. However, this mass loss is slow compared to the short crossing-times of the initially very massive and compact models. This makes sure that the calculated UCDs always survive this period of mass loss. In Fig.~\ref{fig:lagrad}, the time evolution of the Lagrangian radii of one of the models is shown. It can be seen clearly that after an interval of rapid expansion due to the mass loss the UCD-model finally settles back into a new equilibrium. The expansion factor $f_{\rm e}$ of the models is measured by comparing the final ($R_{\rm pl,f}$) with the initial ($R_{\rm pl,0}$) Plummer-radius,
\begin{eqnarray}
  \label{eq:expand}
  f_{\rm e} & = & \frac{R_{\rm pl,f}}{R_{\rm pl,0}}.
\end{eqnarray}
The Plummer-radii are found by fitting Plummer-models to the surface density profiles of the UCDs, using a non-linear least-squares Marquardt-Levenberg algorithm. The Plummer-radius is also identical to the projected half-light radius for the models (see equation~8.57 in \citealt{Kro2008}). Fig.~\ref{fig:exph1s1} shows the expansion factors for all UCD-models with SFE=1 and HE=1. It is visible that among the clusters with top-heavy IMFs (IMFs~1 to~4) the UCD-models with the highest mass expand the least. This is because the more massive UCD-models have shorter crossing times and are therefore closer to the regime of adiabatic mass loss.

Table~\ref{tab:h1s1} shows the final quantities for the models with SFE=1 and HE=1. The models that are the best representations of present-day UCDs at the end of the calculation are marked with a '(+)' in front of the first column. Note that some, but not all of these have the canonical high-mass IMF slope.

We note that assuming HE=1 is not decisive for most of the UCD-models with SFE=1. For instance, if HE=0.03 is assumed for the UCD-models with an initial stellar mass, $M_{*,0}$, of $10^7 \, \rm{M}_{\odot}$ and SFE=1, the mass-loss histories of such models are the same as in the case of HE=1. The same is true for UCD-models with $10^8 \, \rm{M}_{\odot}$ SFE=1 and HE=0.03 if their IMF is the canonical one. On the other hand, for the two most top-heavy IMFs in Tab.~\ref{tab:IMF} (IMFs~1 and~2, $\alpha=1.1$), the UCD-models with $10^8 \, \rm{M}_{\odot}$ and SFE=1 retain most of the gas released by the evolution of massive if HE=0.03, while they are gas-free at all times if HE=1. In the case of H=0.03, these models are very similar to the UCD-models with $10^8 \, \rm{M}_{\odot}$, SFE=0.4 and HE=0.03, which are discussed in Section~\ref{sec:h003s04}.

Thus, it is UCD-models with the most top-heavy IMFs that retain gas the easiest. This is because by assuming that the amount of energy released by a SN does not depend on the mass of the progenitor star (as done in this paper), the total mass set free by the SNe increases more quickly than the total energy provided by the SNe. Also, the luminosity of the stellar population of the UCD, which is the other energy source that powers its mass loss, decreases more rapidly with time for more top-heavy IMFs.

\subsection{SFE=0.4 and HE=1}
\label{sec:h1s04}

\begin{figure}
\centering
\includegraphics[scale=0.80]{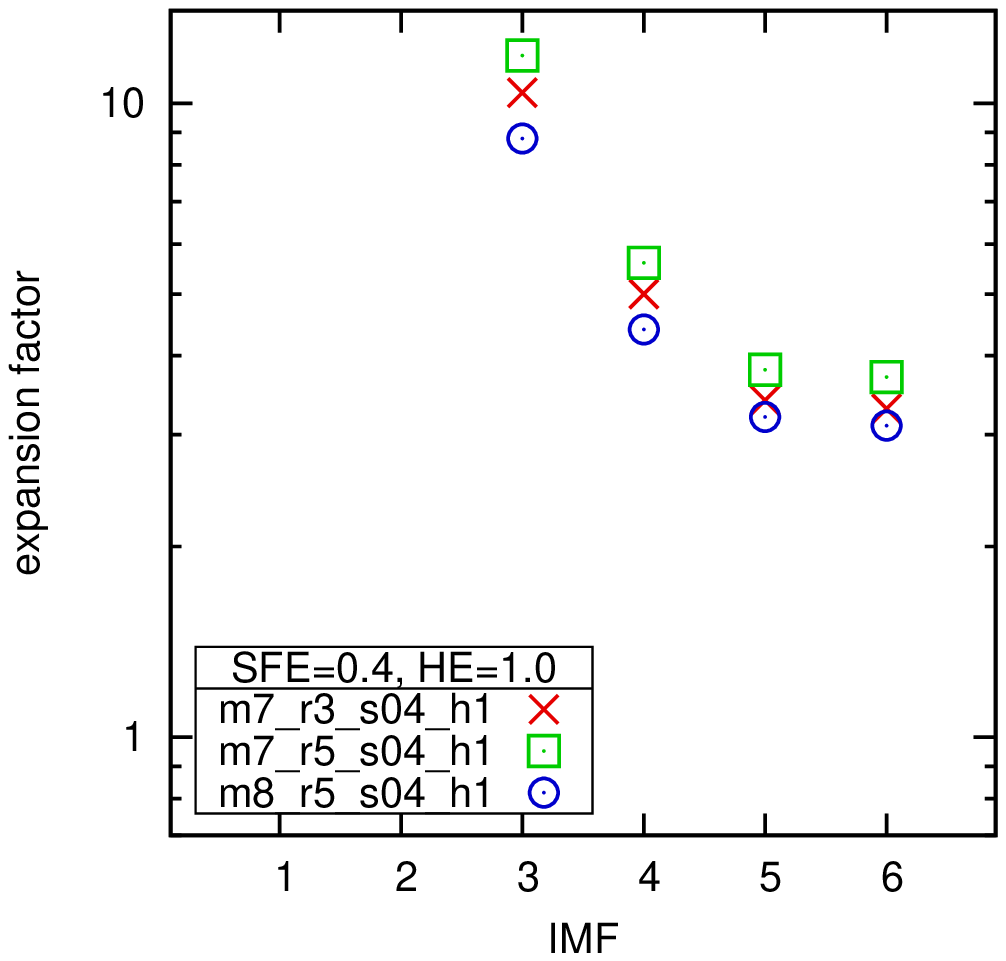}
\caption{Expansion factors $f_{\rm e}$ of the models with a star formation efficiency of 0.4 and a heating efficiency of 1 plotted against the number assigned to the assumed IMF (as in Table~\ref{tab:IMF}). The symbols show the different inital structural parameters of the UCD-models: (red) crosses for $R_{\rm pl,0}=3 \, \rm{pc}$ and $M_{\rm pl,0}=2.5\times 10^7 \, \rm{M}_{\odot}$, (green) squares for $R_{\rm pl,0}=5 \, \rm{pc}$ and $M_{\rm pl,0}=2.5\times 10^7 \, \rm{M}_{\odot}$ and (blue) circles for $R_{\rm pl,0}=5 \, \rm{pc}$ and $M_{\rm pl,0}=2.5\times 10^7 \, \rm{M}_{\odot}$. The modelled UCDs with IMFs~1 and~2 dissolve completely due to their heavy mass loss.}
\label{fig:exph1s04}
\end{figure}

\begin{figure}
  \centering
  \epsfxsize=8cm
  \epsfysize=8cm
  \epsffile{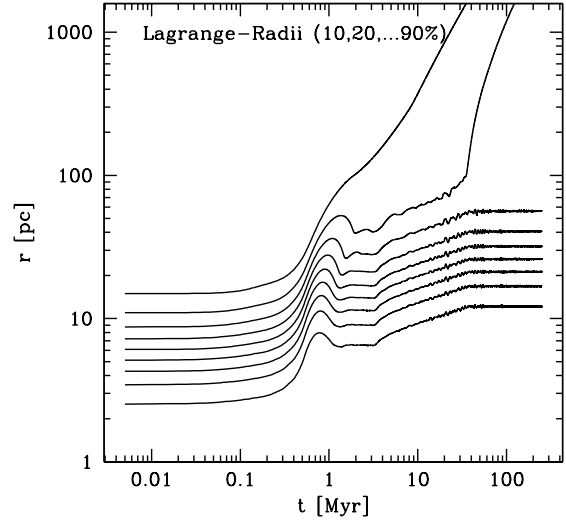}
  \caption{As Fig.~\ref{fig:lagrad}, but for model m8\_r5\_s04\_h1 with IMF~4 ($m_{\rm{max}}=100 \, \rm{M}_{\odot}$, $\alpha=1.9$).}
  \label{fig:h1}
\end{figure}

\begin{table*}
  \centering
  \caption{As Table~\ref{tab:h1s1}, but for a SFE of 0.4 and a HE of 1. These models predict the complete dissolution of the UCD if IMF~1 or IMF~2 are assumed (the ones with high-mass slope $\alpha=1.1$).}
  \label{tab:h1s04}
  \begin{tabular}{lllllllllllllll} \hline
    & model & IMF & $M_{*,\rm{f}}/M_{*,0}$ & $M_{*\rm{b,f}}/M_{*,f}$ & $M_{\rm{f}}$ & $R_{50\rm{,f}}$ & $R_{\rm pl,f}$ & error & $f_{\rm e}$ & $\Sigma_{0,\rm{f}}$ & error & $\sigma_{0,\rm{f}}$ & error \\
    & & & & & [$10^6 \rm{M}_{\odot}$] & [pc] & \multicolumn{2}{c}{[pc]} & & \multicolumn{2}{c}{[M$_{\odot}$\,pc$^{-2}$]} & \multicolumn{2}{c}{[km\,s$^{-1}$]} \\ \hline
    & m7\_r3\_s04\_h1 & 1 & 0.079 & \multicolumn{10}{l}{UCD dissolves completely} \\
    & m7\_r5\_s04\_h1 & 1 & 0.079 & \multicolumn{10}{l}{UCD dissolves completely} \\
    & m8\_r5\_s04\_h1 & 1 & 0.079 & \multicolumn{10}{l}{UCD dissolves completely} \\
    \hline
    & m7\_r3\_s04\_h1 & 2 & 0.114 & \multicolumn{10}{l}{UCD dissolves completely} \\
    & m7\_r5\_s04\_h1 & 2 & 0.114 & \multicolumn{10}{l}{UCD dissolves completely} \\
    & m8\_r5\_s04\_h1 & 2 & 0.114 & \multicolumn{10}{l}{UCD dissolves completely} \\
    \hline
    & m7\_r3\_s04\_h1 & 3 & 0.281 & 0.236 & 0.66 & 30.3 & 33.4 & 0.5 & 10.4 & 298 & 2 & 8.27 & 0.16 \\
    & m7\_r5\_s04\_h1 & 3 & 0.281 & 0.125 & 0.35 & 51.3 & 60.8 & 1.9 & 11.9 & 53 & 1 & 4.49 & 0.06 \\
    & m8\_r5\_s04\_h1 & 3 & 0.281 & 0.482 & 13.5 & 42.9 & 46.7 & 0.6 & 8.8 & 3115 & 16 & 30.95 & 0.30 \\
    \hline
    & m7\_r3\_s04\_h1 & 4 & 0.547 & 0.460 & 2.52 & 19.0 & 16.0 & 0.1 & 5.0 & 3615 & 11 & 19.89 & 0.47 \\
    & m7\_r5\_s04\_h1 & 4 & 0.547 & 0.234 & 1.28 & 28.8 & 28.6 & 0.4 & 5.6 & 672 & 4 & 11.41 & 0.22 \\
    (+) & m8\_r5\_s04\_h1 & 4 & 0.547 & 0.784 & 26.4 & 25.6 & 23.3 & 0.1 & 4.4 & 32182 & 33 & 70.43 & 0.88 \\
    \hline
    & m7\_r3\_s04\_h1 & 5 & 0.770 & 0.626 & 4.82 & 13.6 & 10.9 & 0.1 & 3.4 & 14495 & 35 & 33.46 & 0.50 \\
    (+) & m7\_r5\_s04\_h1 & 5 & 0.770 & 0.300 & 2.31 & 21.4 & 19.2 & 0.1 & 3.8 & 2429 & 8 & 18.22 & 0.51 \\
    & m8\_r5\_s04\_h1 & 5 & 0.770 & 0.896 & 69.0 & 19.2 & 16.7 & 0.0 & 3.2 & 96070 & 120 & 102.64 & 1.87 \\
    \hline
    & m7\_r3\_s04\_h1 & 6 & 0.787 & 0.671 & 5.28 & 13.2 & 10.5 & 0.0 & 3.3 & 16898 & 32 & 35.52 & 1.44 \\
    (+) & m7\_r5\_s04\_h1 & 6 & 0.787 & 0.322 & 2.53 & 21.3 & 19.0 & 0.2 & 3.7 & 2723 & 11 & 18.97 & 0.54 \\
    & m8\_r5\_s04\_h1 & 6 & 0.787 & 0.911 & 71.7 & 18.9 & 16.4 & 0.0 & 3.1 & 103396 & 80 & 105.38 & 1.94 \\
    \hline

  \end{tabular}
\end{table*}

The assumptions SFE=0.4 and HE=1 imply an even more dramatic mass loss than the case of SFE=1 and HE=1 (Section \ref{sec:h1s1}). The energy input of massive stars is high enough to clear the UCD-models of the primordial gas either well before or at the time the first stars end their evolution on the main-sequence. The mass loss is in fact so rapid that a significant fraction of the stars of the UCD-models become unbound, even if an IMF with the canonical high-mass slope ($\alpha=2.3$, IMFs~5 and~6) is assumed. The calculated UCD-models dissolve completely if the IMFs with the flattest high-mass IMF-slopes ($\alpha=1.1$, IMFs~1 and~2) are assumed. Note that they are dissolved by the \emph{combination} of the very rapid expulsion of the primordial gas and the more gentle mass loss though stellar evolution, since an instantaneous loss of 60\% of the initial mass would still leave a bound remnant \citep{Boi2003}, as would the mass loss through stellar evolution alone (cf. Section~\ref{sec:h1s1}).

Analogous to Section~\ref{sec:h1s1}, the expansion of the UCD-models is measured by the ratio between their final Plummer-radii and their initial Plummer-radii and the results are plotted in Fig.~\ref{fig:exph1s04}. It turns out that in this set of models, the UCDs that expand the most are always the ones with the longest crossing times while the UCDs that expand the least are always the ones with the highest initial mass.

{Table~\ref{tab:h1s04} shows the final quantities for the models with SFE=0.4 and HE=1. Model m8\_r5\_s04\_h1 with IMF 4 ($\alpha=1.9$) is the only one of them with a top-heavy IMF and with a good agreement between its final parameters and the parameters observed in UCDs. The evolution of its Lagrange-radii is shown in Fig.~\ref{fig:h1}.}

\subsection{SFE=0.4 and HE=0.03}
\label{sec:h003s04}

\begin{figure}
\centering
\includegraphics[scale=0.80]{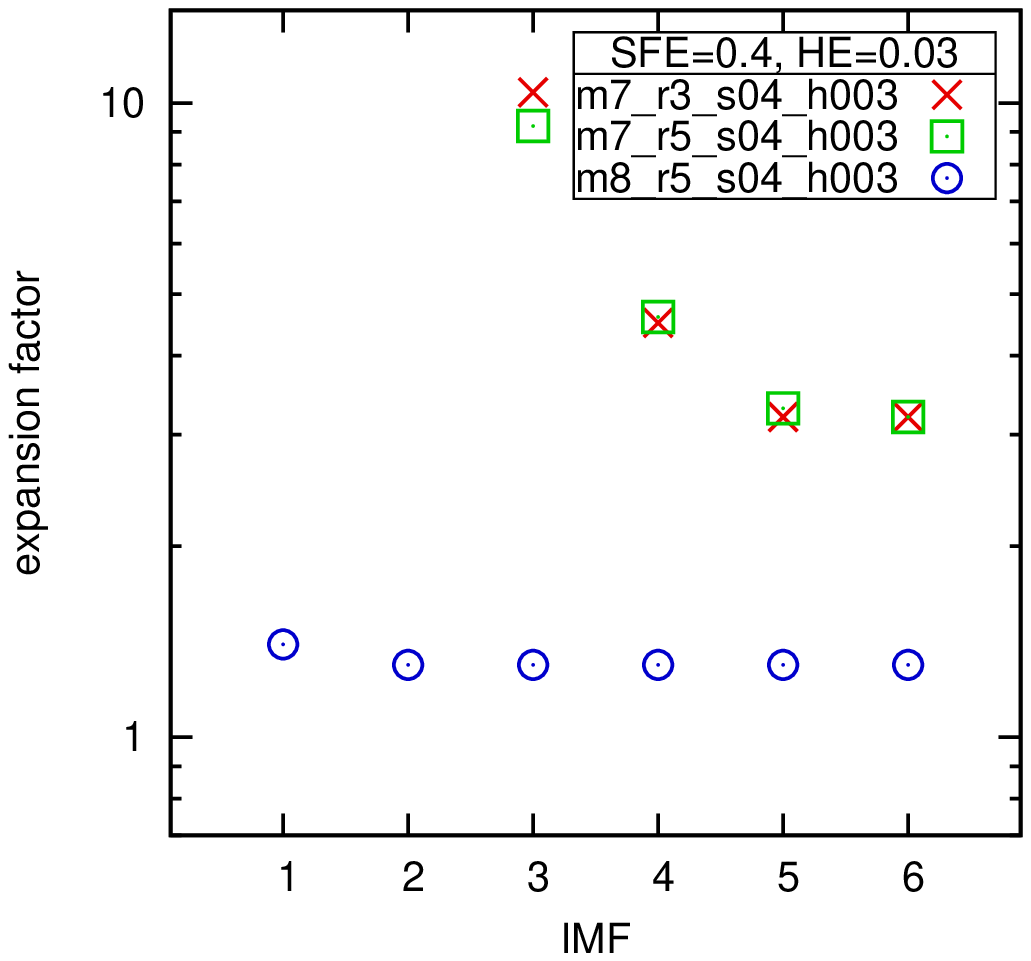}
\caption{As Fig.~\ref{fig:exph1s04}, but for a heating efficiency of 0.03 instead of 1. The modelled UCDs with $M_{*,0}=10^7 \, \rm{M}_{\odot}$ and IMFs~1 or~2 dissolve completely due to their heavy mass loss and are therefore not shown here.}
\label{fig:exph003s04}
\end{figure}

\begin{figure}
  \centering
  \epsfxsize=8cm
  \epsfysize=8cm
  \epsffile{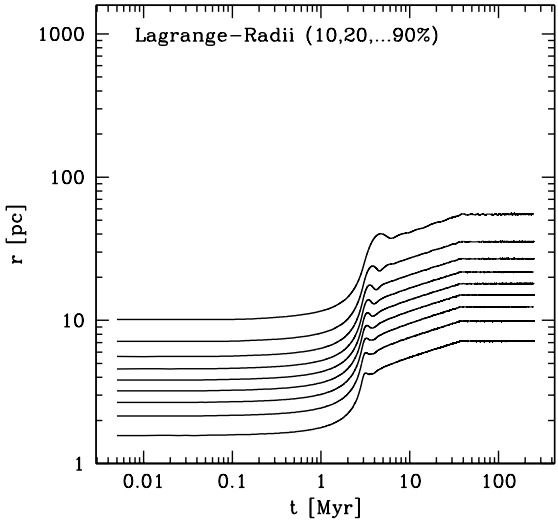}
  \caption{As Fig.~\ref{fig:lagrad}, but for model m7\_r3\_s04\_h003 with IMF~4 ($m_{\rm{max}}=100 \, \rm{M}_{\odot}$, $\alpha=1.9$).}
  \label{fig:h003}
\end{figure}

\begin{figure}
  \centering
  \epsfxsize=8cm
  \epsfysize=8cm
  \epsffile{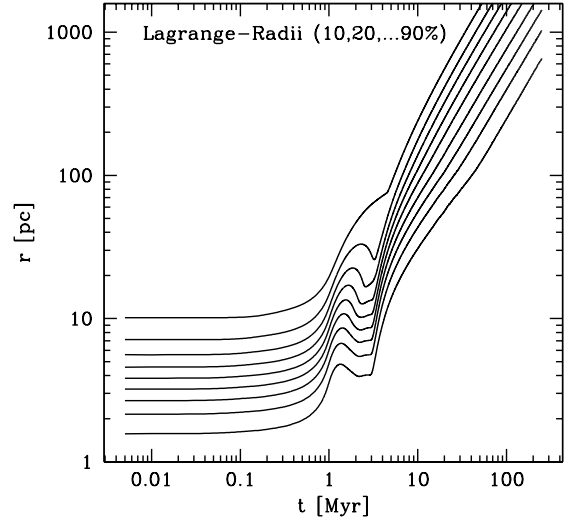}
  \caption{Change of the Lagrangian radii (10, 20, ... 90~\% mass) with time for model m7\_r3\_s04\_h003 with IMF~1 ($m_{\rm{max}}=150 \, \rm{M}_{\odot}$, $\alpha=1.1$). Here the UCD is disrupted by its mass loss, unlike the models shown in Figures.~\ref{fig:lagrad},~\ref{fig:h1} and~\ref{fig:h003}.}
  \label{fig:dissolve}
\end{figure}

\begin{table*}
  \centering
  \caption{As table \ref{tab:h1s1}, but for a SFE of 0.4 and a HE of 0.03. These models predict the complete dissolution of the UCD if IMF~1 or IMF~2 and an initial total mass of $2.5\times10^7 \, \rm{M}_{\odot}$ are assumed. The models with an initial total mass of $2.5\times10^8 \, \rm{M}_{\odot}$ on the other hand keep most of their primordial gas and therefore their evolution is completely different.}
  \label{tab:h003s04}
  \begin{tabular}{lllllllllllllll} \hline
    & model & IMF & $M_{*,\rm{f}}/M_{*,0}$ & $M_{*\rm{b,f}}/M_{*,f}$ & $M_{\rm{f}}$ & $R_{50\rm{,f}}$ & $R_{\rm pl,f}$ & error & $f_{\rm e}$ & $\Sigma_{0,\rm{f}}$ & error & $\sigma_{0,\rm{f}}$ & error \\
    & & & & & [$10^6 \rm{M}_{\odot}$] & [pc] & \multicolumn{2}{c}{[pc]} & & \multicolumn{2}{c}{[M$_{\odot}$\,pc$^{-2}$]} & \multicolumn{2}{c}{[km\,s$^{-1}$]} \\ \hline
    & m7\_r3\_s04\_h003 & 1 & 0.079 & \multicolumn{10}{l}{UCD dissolves completely} \\
    & m7\_r5\_s04\_h003 & 1 & 0.079 & \multicolumn{10}{l}{UCD dissolves completely} \\
    & m8\_r5\_s04\_h003 & 1 & 0.079 & 0.997 & 188.8 & 8.8   & 7.2    & 0.0 & 1.4   & 55622  & 65   & 271.1    & 20.9 \\
    \hline
    & m7\_r3\_s04\_h003 & 2 & 0.114 & \multicolumn{10}{l}{UCD dissolves completely} \\
    & m7\_r5\_s04\_h003 & 2 & 0.114 & \multicolumn{10}{l}{UCD dissolves completely} \\
    & m8\_r5\_s04\_h003 & 2 & 0.114 & 0.997 & 196.5 & 8.5   & 7.0    & 0.0 & 1.3  & 86030   & 150 & 282.80 & 23.80 \\
    \hline
    & m7\_r3\_s04\_h003 & 3 & 0.281 & 0.742 & 2.09   & 29.0 & 28.1 & 0.2 & 10.4 & 1164     & 3     & 14.64    & 0.17 \\
    & m7\_r5\_s04\_h003 & 3 & 0.281 & 0.469 & 1.32   & 42.5 & 46.8 & 0.6 & 9.2   & 308        & 1     & 9.69      & 0.11 \\
    & m8\_r5\_s04\_h003 & 3 & 0.281 & 0.997 & 200.0 & 8.4   & 6.9   & 0.0 & 1.3   & 215810 & 340 & 287.40 & 24.70 \\
    \hline
    (+) & m7\_r3\_s04\_h003 & 4 & 0.547 & 0.944 & 5.16   & 17.2 & 14.5 & 0.0 & 4.5 & 9278     & 9      & 29.66   & 0.43 \\
    & m7\_r5\_s04\_h003       & 4 & 0.547 & 0.835 & 4.57   & 25.9 & 23.3 & 0.0 & 4.6 & 3376     & 6      & 22.73   & 0.30 \\
    & m8\_r5\_s04\_h003       & 4 & 0.547 & 0.996 & 202.4 & 8.4   & 6.8   & 0.0 & 1.3 & 424440 & 500 & 290.70 & 25.60 \\
    \hline
    (+) & m7\_r3\_s04\_h003 & 5 & 0.770 & 0.978 & 7.53   & 12.4 & 10.3 & 0.0 & 3.2 & 26578    & 18   & 42.36   & 0.79 \\
    (+) & m7\_r5\_s04\_h003 & 5 & 0.770 & 0.906 & 6.98   & 18.9 & 16.6 & 0.0 & 3.3 & 9973      & 140 & 33.06   & 0.58 \\
    & m8\_r5\_s04\_h003 	   & 5 & 0.770 & 0.996 & 208.4 & 8.2   & 6.7    & 0.0 & 1.3 & 623290 & 640 & 299.50 & 27.70 \\
    \hline
    (+) & m7\_r3\_s04\_h003 & 6 & 0.787 & 0.980 & 7.71   & 12.2 & 10.1 & 0.0 & 3.2 & 28116   & 25 & 43.33   & 0.82 \\
    (+) & m7\_r5\_s04\_h003 & 6 & 0.787 & 0.919 & 7.23   & 18.7 & 16.2 & 0.0 & 3.2 & 10697   & 17 & 33.76    & 0.58 \\
    & m8\_r5\_s04\_h003 	   & 6 & 0.787 & 0.996 & 208.4 & 8.2   & 6.7    & 0.0 & 1.3 & 633900 & 16 & 298.90 & 27.00 \\
    \hline
  \end{tabular}
\end{table*}

In the case of a moderately high star formation efficiency (SFE=0.4) and and low heating efficiency (HE=0.03), the UCD-models with $M_{\rm pl,0}=2.5\times 10^7 \, \rm{M}_{\odot}$ are gas-free at the end of the computation (as are all models with HE=1). In contrast to that, the models with $M_{\rm pl,0}=2.5\times 10^8 \, \rm{M}_{\odot}$ keep most of the gas at such a low HE, so that these UCD-models are predicted to expand barely and to consist mainly of gas at the end of the integration, implying very high $M/L_V$ ratios. (In the two most extreme cases, where the high-mass IMF slope is $\alpha=1.1$, approximately 5\% of the total mass of the UCD-model is stars at that time, while the rest is gas.) This is a very implausible situation. It is more likely that if the heating efficiency is too low to drive the gas out of the cluster, the star formation efficiency would become higher through new star formation episodes, until eventually all matter is locked up in low-mass stars and thereby a SFE of 1 is approached. 

However, the half-mass radius of a UCD would hardly change with time in this case, since it keeps most of its initial mass (cf. equation~\ref{adiabatic}), while the initial half-mass radii of UCDs suggested in this paper are clearly smaller than the half-mass radii of observed present-day UCDs (cf. table~5 in \citealt{Mie2008b}). This means that, if a UCD indeed retains most of its mass, it must be born with an initial half-mass radius close to the observed values. By calculating mass-loss histories of UCD-models with $M_{\rm pl,0}=2.5\times 10^8 \, \rm{M}_{\odot}$, SFE=0.4 and HE=0.03 for different initial Plummer radii (by the method described in Section~\ref{sec:algorithm}), it turns out  that the energy input from massive stars is sufficient to remove all gas from these UCD-models at an initial projected half-mass radius (i.e. Plummer radius) of 12 pc instead of 5 pc, even if the UCD-models have the canonical IMF. In contrast to that, the most massive observed UCDs, with masses of $\approx 10^8 \, \rm{M}_{\odot}$, are reported to have half-mass radii of $\approx 100$ pc. Thus, adopting a major star burst as the scenario for the birth of a UCD (as done in this paper), an object that is able to keep gas after the star burst would be too compact to evolve into an UCD. On the other hand, if the object has an initial half-mass radius that allows it to evolve into a UCD, it would loose its gas on a time scale of a few Myr. This excludes the above scenario where a UCD forms a substantial part of its stellar population over a longer period of time after the initial star burst.

Fig.~\ref{fig:exph003s04} depicts the expansion rates of the UCDs with SFE=0.4 and HE=0.03 and shows a strong difference between the modelled UCDs with $M_{\rm pl,0}=2.5\times 10^8 \, \rm{M}_{\odot}$ and the ones with $M_{\rm pl,0}=2.5\times 10^7 \, \rm{M}_{\odot}$. While the less massive UCD-models expand almost as much as in the case of HE=1 (see Sections~\ref{sec:h1s1} and~\ref{sec:h1s04}), the extension of the more massive UCD-models hardly changes.

The final quantities found for the models with SFE=0.4 and HE=0.03 are shown in Table~\ref{tab:h003s04}. Except for models with IMFs that have the canonical high-mass slope, only model m7\_r3\_s04\_h003 with IMF~4 ($\alpha=1.9$) is a good representation of a UCD. The time-evolution of its Lagrange-radii is shown in Fig.~\ref{fig:h003}. Fig.~\ref{fig:dissolve} on the other hand illustrates the evolution of a UCD that dissolves completely because of extreme mass loss due to its very top-heavy IMF. The only difference to the model shown in Fig.~\ref{fig:h003} is that IMF~1($\alpha=1.1$) instead of IMF~4 was assumed.

\subsection{Implications on the initial parameters of UCDs}
\label{sec:implications}

\begin{figure}
\centering
\includegraphics[scale=0.80]{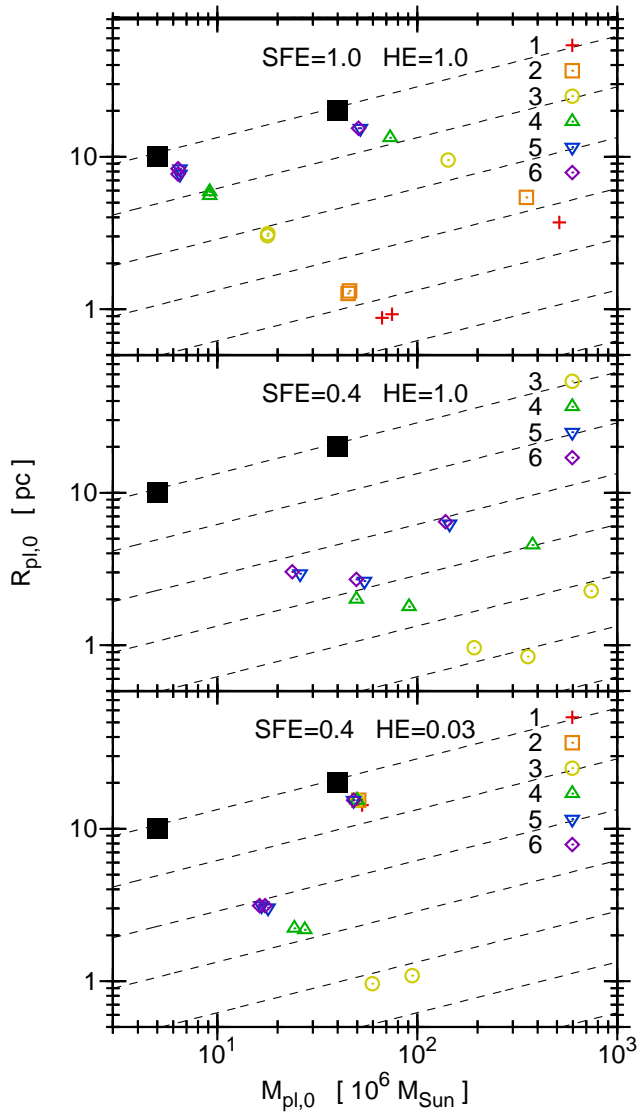}
\caption{Estimated initial masses and Plummer-radii that would lead to representative, UCD-type objects. The black square to the left represents in each panel an observed typical small UCD with a mass of $5\times10^6 \, \rm{M}_{\odot}$ and a Plummer radius of 10 pc, whereas the black square to the right represents an observed typical large UCD with a mass of $40\times10^6 \, \rm{M}_{\odot}$ and a Plummer radius of 20 pc. The remaining symbols show estimated initial masses and Plummer-radii, that would lead to one of these two representative UCDs with the IMFs from Table~\ref{tab:IMF}, identified here by the number assigned to them in that table. The assumed star formation efficiency and heating efficiency are indicated at the top of each panel. For the more massive UCD-like object, the estimated initial parameters are based on the total mass loss (through stellar and dynamical evolution) and expansion factors of the models starting with a total stellar initial mass $M_{*,0}=10^8 \, \rm{M}_{\odot}$ (Table~\ref{tab:ini}), while the estimates for the less massive UCD-like object are based on the models with $M_{*,0}=10^7 \, \rm{M}_{\odot}$ (Table~\ref{tab:ini}). The dashed lines in each panel show constant central densities, starting from $10^3 \, \rm{M}_{\odot} \, \rm{pc}^{-3}$ and increasing by a factor of ten downward with each line. Note that the initial conditions resulting in the more massive representative UCD in the lowermost panel (SFE=0.4 and HE=0.03) are based on models where hardly any gas is lost from the UCD-model while no more stars are formed from this material, which is a unrealistic scenario (see Section~\ref{sec:h003s04}).}
\label{fig:Inipar}
\end{figure}

\begin{figure}
\centering
\includegraphics[scale=0.80]{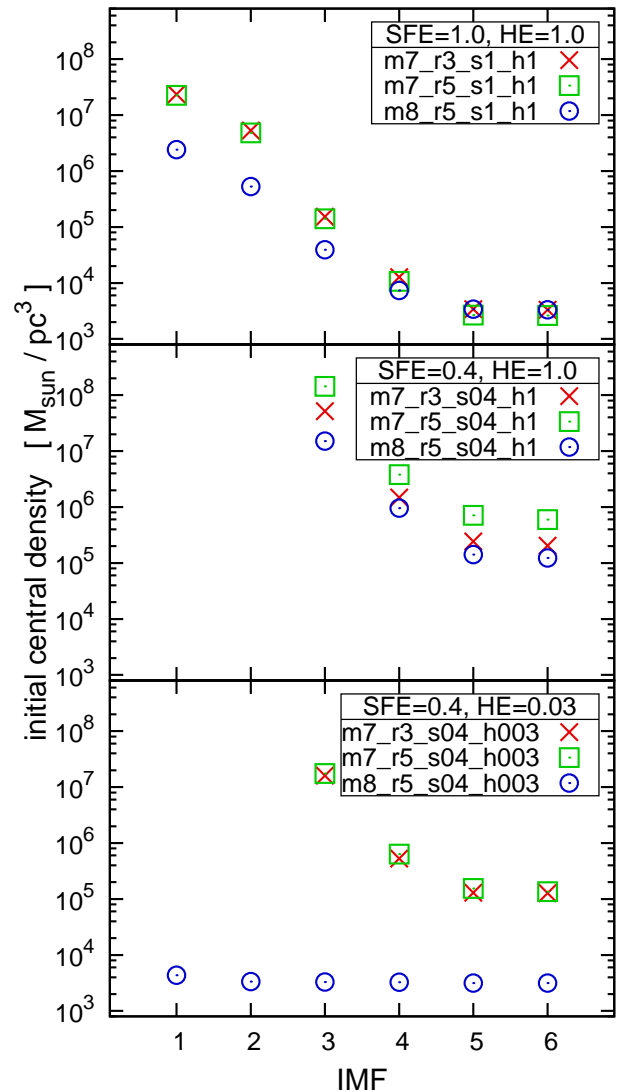}
\caption{The initial central densities that would lead to the representative UCD-type objects shown as black squares in Fig.~\ref{fig:Inipar}, given the mass losses and expansion factors of the UCD-models calculated in this paper for the IMFs listed in Table~\ref{tab:IMF}. It is assumed (as in Fig.~\ref{fig:Inipar}) that the mass loss and the expansion experienced by an object that evolves into the more massive representative UCD (with a mass of $40\times10^6 \, \rm{M}_{\odot}$ and a Plummer radius of 20 pc) is given by the mass losses and expansions calculated for the UCD-models that start with an initial total stellar mass of $10^8 \, \rm{M}_{\odot}$ (Table~\ref{tab:ini}). The evolution of the less massive representative UCD (with a mass of $5\times10^6 \, \rm{M}_{\odot}$ and a Plummer radius of 10 pc) from its initial state is thought to be consistent with the mass losses and expansion factors found for the UCD-models starting with an initial total stellar mass of $10^7 \, \rm{M}_{\odot}$ (Table~\ref{tab:ini}). The central densities of the initial states of UCDs in this figure are thus given by the initial masses and Plummer radii assigned to them in Fig.~(\ref{fig:Inipar}), using equation~(\ref{eq:rhoPlummer}) with $R=0$.}
\label{fig:Densities}
\end{figure}

Based on the fraction of the mass that is lost from the modeled stellar systems and the factors by which they expand due to mass loss, initial conditions that would lead to UCD-like objects can be estimated. The results of such estimates are shown in Fig.~\ref{fig:Inipar}. Thus, UCDs may have been born from extremely compact configurations with densities ranging up to $10^8 \rm{M}_{\odot} \, \rm{pc}^{-3}$. These numbers have admittedly to be taken with caution, since the expansion factors and the total mass loss of the objects have been derived for stellar systems with different initial parameters, using mass loss histories through stellar processes that were specifically created for them. Note however the similarity between the expansion factors of models with the same initial mass and IMF, but different initial radii (Figs.~\ref{fig:exph1s1},~\ref{fig:exph1s04} and~\ref{fig:exph003s04}). Analogous calculations to the ones performed here, but with the initial parameters plotted in Fig.~\ref{fig:Inipar} are therefore likely to lead to final parameters that represent the actual parameters of UCDs better, but this needs to be studied in follow-up work.

The initial central densities following from the pairs of initial masses and initial Plummer-radii plotted in Fig.~\ref{fig:Inipar} are shown in Fig.~\ref{fig:Densities}. The initial parameters that would lead to UCD-type objects according to Figs.~\ref{fig:Inipar} and~\ref{fig:Densities} can be compared to the initial parameters of the UCD-models listed in Tab~\ref{tab:ini}, whose early evolution was calculated in this paper. It thereby becomes apparent that the initial conditions resulting in UCDs may be even more extreme than the ones that are specified in Table~\ref{tab:ini}. Thus, encounters of proto-stars with stars, as discussed in Section~\ref{sec:encounters}, may be even more relevant for the star formation in actual UCDs than for the UCD-models calculated in this paper. The UCDs may even have been dense enough for frequent collisions between stars, so that this process could also have shaped their IMF (cf. \citealt{Bon1998}).

Note the similarity of Figs.~(\ref{fig:exph1s1})~(\ref{fig:exph1s04}) and (\ref{fig:exph003s04}) with the corresponding panels of Fig.~(\ref{fig:Densities}), except for the different scaling. This is because the expansion factor enters with the third power into the calculation of the initial density for a given final mass and final Plummer-radius according to equation~(\ref{eq:rhoPlummer}), while the dependency on the lost mass is only linear.

Also note that the negligence of compact remnants induces a bias on the estimated initial parameters: Remnants kept by the UCD-model diminish the mass that leaves the UCD-model and thereby also reduce its expansion. For arriving at the mass and the Plummer radius of the representative UCD-type objects plotted in Fig.~\ref{fig:Inipar}, a UCD-model that keeps some of the mass of the massive stars in the form of remnants would thus need a larger initial radius and a smaller initial mass than a UCD-model that looses all remnants from massive stars. Consequently, the initial density of the UCD-model that keeps some remnants would also be smaller. The total mass of the remnants remaining in the UCDs has however been argued unlikely to be much larger than 10 per cent of the total mass of their progenitor stars (Section~\ref{sec:remnants}) and the bias on the initial parameters shown in Fig.~\ref{fig:Inipar} would be of the same order. The large implied mass-loss through the evolution of massive stars does not contradict the high $M/L_V$ ratios of UCDs, if the number of massive stars was sufficient in them, i.e. their IMF was top-heavy enough (cf. DKB).

The bias caused by the the negligence of compact remnants may however be alleviated by an opposed bias. This opposed bias comes from the fact that non-adiabatic behaviour of the UCD-models was taken into account in the actual calculation of their dynamical behaviour, but not in the modelling of the mass loss driving the early evolution of the UCD-models (see Section~\ref{sec:algorithm}).

\section{Summary and conclusions}
\label{sec:conc}

\begin{figure}
\centering
\includegraphics[scale=0.80]{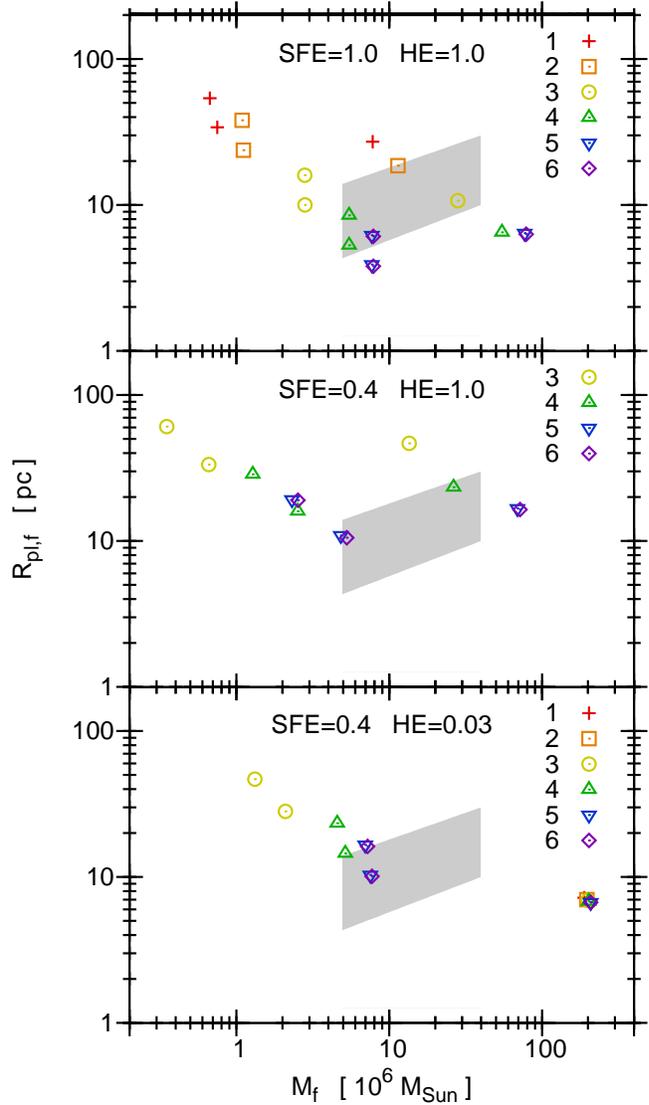}
\caption{The final Plummer-radii against their final masses of all models that have not dissolved at the end of the integration. Different symbols encode different IMFs, identified by the numbers assigned to them in Tab~\ref{tab:IMF}.  The assumed star formation efficiency and heating efficiency is given at the top of each panel. The shaded regions indicate the parameter space occupied by real UCDs.}
\label{fig:Finpar}
\end{figure}

\begin{table*}
\caption{Consistency-check between the models for the remnant populations of the UCDs discussed in DKB based on their dynamical $M/L_V$ ratios with the models discussed here for their early dynamical evolution, assuming SFE=1 and HE=1 (for other SFEs and HEs see Section~\ref{sec:conc}). DKB consider two different ages for the UCDs. The first column specifies various remnant populations of the UCDs, as found in DKB. They differ by the mass of the SN remnants and which fraction of them remains bound to its host UCD. Concerning the kind of compact remnant a SN leaves, it is assumed that stars with initial masses between $8 \, \rm{M}_{\odot}$ and $25 \, \rm{M}_{\odot}$ become neutron stars with a mass of $1.35 \, \rm{M}_{\odot}$. Stars with initial masses above $25 \, \rm{M}_{\odot}$ become black holes with either 10 or 50 per cent of the initial mass of their progenitors ($m_{\rm{BH}}=0.1 \rm{or}\ 0.5 m_*$). The upper mass limit of the IMF is $100 \, \rm{M}_{\odot}$ in all models. The second column displays the high-mass IMF slopes, $\overline{\alpha}$, which correspond to these remnant populations, given the mean dynamical $M/L_V$ ratio of the UCDs (see DKB for details). The uncertainties on $\overline{\alpha}$ are calculated from the uncertainties on the mean $M/L_V$ ratios. Columns~3 to~5 indicate how consistent the models discussed in DKB are with the ones discussed here. In this context, a '$+$' means that $\overline{\alpha} \pm 0.1$ agrees with the high-mass IMF slope in one of the UCD-models marked with a '(+)' in Table~\ref{tab:h1s1}, i.e. $|\overline{\alpha} -\alpha|< 0.1$, where $\alpha$ is the high-mass slope of the IMF whose number according to Table~\ref{tab:IMF} is given in brackets.  A '$\ocircle$' has an analogous meaning, but it is only required that $|\overline{\alpha} -\alpha|< 0.2$. A '$-$' indicates that a model with these initial condidtions can only reproduce final parameters as observed in UCDs with an IMF with $|\overline{\alpha} -\alpha|> 0.2$.}
\centering
\vspace{2mm}
\begin{tabular}{lllll}
\hline
&&&&\\[-10pt]
remnant population 	   & $\overline{\alpha}$ 	     & m7\_r3\_h1\_s1 	      & m7\_r5\_h1\_s1 	     & m8\_r5\_h1\_s1 \\
\hline
&&&&\\
\multicolumn{5}{c}{assumed age of 13 Gyr}\\
&&&&\\
no SN remnants  retained					     			     & $1.35_{-0.17}^{+0.23}$ & $-$ & $-$ & $\ocircle \ (3)$ \\
20 per cent of the SN remnants retained, $m_{\rm{BH}}=0.1 m_*$ & $1.57_{-0.12}^{+0.17}$ & $-$ & $-$ & $+ \ (3)$ \\
20 per cent of the SN remnants retained, $m_{\rm{BH}}=0.5 m_*$ & $1.78_{-0.09}^{+0.13}$ & $\ocircle \ (4)$ & $ \ocircle \ (4)$ & $-$ \\
all SN remnants retained, $m_{\rm{BH}}=0.1 m_*$ 		      & $1.85_{-0.10}^{+0.14}$ & $+ \ (4)$ & $+ \ (4)$ & $-$ \\
&&&&\\
\multicolumn{5}{c}{assumed age of 7 Gyr}\\
&&&&\\
no SN remnants retained 						       		     & $0.49_{-0.08}^{+0.09}$ & $-$ & $-$ & $-$ \\
20 per cent of the SN remnants retained, $m_{\rm{BH}}=0.1 m_*$ & $1.04_{-0.04}^{+0.05}$ & $-$ & $-$ & $+ \ (2)$ \\
20 per cent of the SN remnants retained, $m_{\rm{BH}}=0.5 m_*$ & $1.34_{-0.04}^{+0.04}$ & $-$ & $-$ & $\ocircle \ (3)$ \\
all SN remnants retained, $m_{\rm{BH}}=0.1 m_*$ 	       	     & $1.40_{-0.04}^{+0.05}$ & $-$ & $-$ & $\ocircle \ (3)$ \\
\hline
\end{tabular}
\label{tab:comp}
\end{table*}

We calculate the early evolution of extremely massive star clusters, using the particle-mesh code Superbox. Their initial radii are chosen in concordance with typical values for globular clusters (GCs), while their initial masses reflect the masses of ultra compact dwarf galaxies (UCDs). The early evolution of a star cluster is driven by mass loss through gas expulsion and stellar evolution. This mass loss is treated by reducing the mass of each particle in accordance with previously tabulated mass loss histories, so that the total mass of all particles agrees with the total mass of the UCDs as given in those tables. The rate and the magnitude of the mass loss depends in particular on the stellar initial mass function (IMF). Since it was suggested that UCDs may have formed with a top-heavy IMF (DKB), the integrations use mass loss tables not only for the canonical IMF but also with different top-heavy IMFs. A possible explanation for why the IMF in UCDs could be top-heavy is encounters between proto-stars and stars. If UCDs indeed formed as the most massive star clusters, as suggested in this paper, such encounters would be quite likely in emerging UCDs. In contrast, such encounters are not very probable in stellar systems that evolve into star clusters like the Orion nebula cluster. This implies that star formation in UCDs may be influenced by processes that  do not play a significant role in less massive stellar systems. The final masses and Plummer-radii resulting from the calculations in this paper are shown in Fig.~\ref{fig:Finpar}.

The possible initial conditions we uncover here (Figs.~\ref{fig:Inipar} and~\ref{fig:Densities}) include densities as high as $10^8 \, \rm{M}_{\odot} \, \rm{pc}^{-3}$ for the forming UCDs with top-heavy IMFs ($\alpha \le 1.9$). The super nova rates are at times as high as one per year in the UCD-models with with an initial stellar mass of $10^7 \, \rm{M}_{\odot}$ and the most top-heavy IMFs (Fig.~\ref{fig:SN}) and higher by a factor of 10 in the UCD-models with an initial stellar mass of $10^8 \, \rm{M}_{\odot}$.

Starting from our initial conditions (Table~\ref{tab:ini}), we seek those final models that represent UCDs in terms of their radii, masses and $M/L_V$-ratios in the following.

\begin{itemize}
 \item If the UCDs form as star clusters with a high star-formation efficiency and a high heating efficiency (as discussed with the case SFE=1 and HE=1; see Section~\ref{sec:h1s1}), the properties of present-day UCDs are reproduced from models with all IMFs in Table~\ref{tab:IMF} except IMF~1, i.e. with stellar populations with high-mass IMFs in the whole range from $\alpha=2.3$ (canonical IMF) to $\alpha=1.1$ (see Table~\ref{tab:h1s1} and Fig.~\ref{fig:Finpar}). The different models imply however different ages and different stellar remnant populations for the UCDs, because they are constrained by the average $M/L_V$ ratio that is observed for UCDs (cf. table~3 in DKB). A consistency check between the models in this paper and the models in DKB is provided in Table~\ref{tab:comp}.Note that the model from DKB where stars with an initial mass larger than $25 \, \rm{M}_{\odot}$ are assumed to evolve into black holes that have 50 per cent of the mass of their progenitors and all compact remnants are thought to be retained by the UCD is not listed in Table~\ref{tab:comp}. This is because it is not consistent with the assumption that UCDs loose most of the mass that was initially locked up in their massive stars. However, it seems likely that UCDs loose indeed most of this mass (see Section~\ref{sec:remnants}. The model from DKB where all matter is kept within the UCDs is also omitted from Table~\ref{tab:comp}, even though it would seem consistent with the UCD-models with initial stellar mass of $10^8 \, \rm{M}_{\odot}$, a SFE of 0.4 and a HE of 0.03. These initial parameters lead however to a unrealistic situation at the end of the calculation, because the UCD-models that can evolve this way stay too compact for being consistent with real UCDs (see Section~\ref{sec:h003s04}).
 
As a result of the comparison shown in Table~\ref{tab:comp}, the models with IMFs~5 and~6 (canonical high-mass IMF slope) can be excluded as formation scenarios for the UCDs as a class of objects. This is because these models suggest that the $M/L_V$ ratio is consistent with the ones predicted by  simple stellar population models, which is not the case for UCDs \citep{Dab2008,Mie2008b}. The models with IMFs~5 and~6 would however be consistent with the UCDs in the Fornax Cluster if they are very old, because their average $M/L_V$ ratio is somewhat lower than the ones of UCDs in general \citep{Mie2008b}.

  \item If the UCDs form as star clusters with a moderate star-formation efficiency and high heating efficiency (as discussed with the case SFE=0.4 and HE=1; see Section~\ref{sec:h1s04}), extremely top-heavy IMFs ($\alpha=1.1$) can be excluded because they would lead to the complete dissolution of the cluster. Model m8\_r5\_h1\_s04 with IMF~4 ($\alpha=1.9$) resembles a massive present-day UCD at the end of the integration. A comparison with table~3 in DKB shows that this model is consistent with two cases listed there. The first of them is the case of the UCDs being 13 Gyr old, keeping 20 per cent of the SN remnants and black holes, which retain 50 per cent of the mass of their progenitor stars. The second is the case of the UCDs being 13 Gyr old, keeping all SN remnants and black holes having 10 per cent of the mass of their progenitor stars. These would be the only cases where a table analogous to Table~\ref{tab:comp} would indicate consistency between the UCD-models here and the ones in DKB. As in the case of SFE=1 and HE=1, the models with IMFs~5 and~6 (canonical high-mass IMF slope) can be excluded as formation scenarios for the UCDs as a class of objects, because of their too-low $M/L_V$ ratio \citep{Dab2008,Mie2008b}.
  
 \item If the UCDs form as star clusters with a moderate star-formation efficiency and low heating efficiency (as discussed with the case SFE=0.4 and HE=0.03; see Section~\ref{sec:h003s04}), the models with an initial mass of $2.5\times10^8 \, \rm{M}_{\odot}$ lead to the unrealistic case that gas of the order of $10^8 \, \rm{M}_{\odot}$ is confined on a very small volume at the end of our calculations. Models starting with an initial mass of $2.5\times10^7 \, \rm{M}_{\odot}$ on the other hand dissolve for the two most top-heavy IMFs, like in the case of SFE=0.4 and HE=1. Model m7\_r3\_s04\_h003 with IMF~4 ($\alpha=1.9$) is similar to a small present-day UCD at the end of the calculation.  A comparison with table~3 in DKB shows that this model is consistent with two cases listed there. The first of them is the case of the UCDs being 13 Gyr old, keeping 20 per cent of the SN remnants and black holes, which retain 50 per cent of the mass of their progenitor stars. The second is the case of the UCDs being 13 Gyr old, keeping all SN remnants and black holes having 10 per cent of the mass of their progenitor stars. These would be the only cases where a table analogous to Table~\ref{tab:comp} would indicate consistency between the UCD-models here and the ones in DKB.
 
Note the difference to the UCD-models with moderate SFE and high HE. In the case of a moderate SFE and high HE, consistency between the UCD-models from this paper and the models for the remnant populations of UCDs from DKB is reached for a UCD-model starting with an initial stellar mass of $10^8 \, \rm{M}_{\odot}$, whereas in the case of a moderate SFE and a low HE consistency is reached for a UCD-model starting with an initial stellar mass of $10^7 \, \rm{M}_{\odot}$. The UCD-models from this paper are however in both cases consistent with the same models for the remnant populations of UCDs in DKB. As in the case SFE=1 and HE=1, the models with IMFs~5 and~6 (canonical high-mass IMF slope) can be excluded as formation scenarios for the UCDs as a class of objects, because of their too-low $M/L_V$ ratio \citep{Dab2008,Mie2008b}.
 \end{itemize}
 
Thus, in summary, the preferred solution of the initial conditions problem for a SFE of 0.4 are a proto-UCD with a stellar initial mass of $10^8 \, \rm{M}_{\odot}$ and a projected half-mass radius of 5 pc (HE=1) or a proto-UCD with a stellar initial mass of $10^7 \, \rm{M}_{\odot}$ and a projected half-mass radius of 3 pc (HE=0.03). UCD-models with SFE=1 are discussed in Section~\ref{sec:h1s1} and for SFE=1 and HE=1, the preferred solutions are presented in Table~\ref{tab:comp}.
 
The comparison between the final parameters of our models and observed parameters of present-day UCDs contain some uncertainties for initial parameters that lead to the formation of UCDs because of a number of approximations and simplifying assumptions (also see Section~\ref{sec:setup} for this matter). For instance, the density profiles of UCDs are usually better fitted by a King profile \citep{ki66} than by a Plummer profile. Also, the calculations performed here stop at 250 Myr, whereas UCDs are $\approx 10$ Gyr old. Thus UCDs will have suffered from adiabatic mass loss through the evolution of intermediate-mass stars, if the material expelled by them is not used up in the formation of subsequent stellar populations. Finally, the tidal field of the host galaxy of the UCD may play a role for its evolution on a Gyr time-scale. The performed comparison demonstrates however that also the more rapid early mass loss triggered by an over-abundance of massive stars does not necessarily lead to complete dissolution of massive, dense stellar systems, but can result in objects similar to a UCD. The existence of UCDs is therefore not in contradiction with their formation with a top-heavy IMF. In a number of cases, a top-heavy IMF leads to a strong inflation of the modelled UCDs, but does not completely disintegrate them.

\section*{Acknowledgements}
JD acknowledges support through DFG grant KR1635/13. MF was supported through GEMINI project no. 32070002.

\label{lastpage}

\end{document}